\documentclass[12pt,3p,authoryear]{elsarticle}
\usepackage[T1]{fontenc}
\usepackage{amsmath,amssymb,amsfonts}
\usepackage{natbib}
\defcitealias{NW1987}{Newey-West (1987)}
\defcitealias{FamaMcBeth1973}{Fama-MacBeth (1973)}
\usepackage{graphicx}
\usepackage{epstopdf}
\PassOptionsToPackage{hyphens}{url}
\usepackage{url}
\usepackage[hidelinks]{hyperref} 
\usepackage{booktabs}
\usepackage{xspace}
\usepackage{setspace}
\usepackage{color}
\usepackage{subdepth}
\usepackage{float}
\usepackage[font={small,singlespacing},labelfont={bf,small},skip=0.2cm]{caption}
\floatstyle{plaintop}
\restylefloat{table}
\restylefloat{figure}	
\graphicspath{{./img/}}
\usepackage{multirow}
\usepackage{caption}
\usepackage{subcaption}
\usepackage{makecell}
\usepackage{ragged2e}
\usepackage{pdflscape}
\usepackage{afterpage}
\usepackage{chngcntr}
\usepackage{placeins}
		
\journal{arXiv}

\newcommand{\ie}{\emph{i.e.}\xspace}
\newcommand{\eg}{\emph{e.g.}\xspace}

\newcommand{\insertfloat}[1]{%
\begin{center}
[Insert~#1 about here.]%
\end{center}%
}

\begin{document}
\begin{frontmatter}

\title{Revisiting \citet{Boehmeretal2021}:\\ Recent Period, Alternative Method, Different Conclusions\tnoteref{label1}} 
\tnotetext[label1]{We thank Mehmet Saglam for useful comments. We are grateful to the Natural Sciences and Engineering Research Council of Canada (grant
RGPIN-2022-03767) and the Group for Research in Decision Analysis (GERAD) for their financial support.}
\author[gerad]{David Ardia}
\ead{david.ardia@hec.ca}
\author[gerad,fin]{Clément Aymard\corref{cor1}}
\ead{clement.aymard@hec.ca}
\cortext[cor1]{Corresponding author. HEC Montréal, 3000 Chemin de la Côte-Sainte-Catherine, Montreal, QC H3T 2A7. Phone: +1 514 340 6103.}
\author[fin]{Tolga Cenesizoglu}
\ead{tolga.cenesizoglu@hec.ca}
\address[gerad]{GERAD \& Department of Decision Sciences, HEC Montréal, Montréal, Canada}
\address[fin]{Department of Finance, HEC Montréal, Montréal, Canada}

\begin{abstract}
We reassess \citet[BJZZ]{Boehmeretal2021}'s seminal work on the predictive power of retail order imbalance (ROI) for future stock returns. First, we replicate their 2010-2015 analysis in the more recent 2016-2021 period. We find that the ROI's predictive power weakens significantly. Specifically, past ROI can no longer predict weekly returns on large-cap stocks, and the long-short strategy based on past ROI is no longer profitable. Second, we analyze the effect of using the alternative quote midpoint (QMP) method to identify and sign retail trades on their main conclusions. While the results based on the QMP method align with BJZZ's findings in 2010-2015, the two methods provide different conclusions in 2016-2021. Our study shows that BJZZ's original findings are sensitive to the sample period and the approach to identify ROIs.
\end{abstract}
\end{frontmatter}

\doublespacing

\newpage
\section{Introduction}
\noindent 
A central question in the literature on retail investors, as succinctly expressed by \citet[BJZZ]{Boehmeretal2021} in their opening sentence, is: ``\textit{Can retail equity investors predict future stock returns, or do they make systematic, costly mistakes in their trading decisions?}'' While earlier studies, such as \citet{BarberOdean2000} and \citet{BarberOdean2008}, did not find significant predictive patterns between retail investors' trading and future returns, more recent research suggests that retail investors' order flow has predictive power for future returns \citep[\eg,][]{KanielSaarTitman2008, BarberOdeanZhu2009, Kanieletal2012,KelleyTetlock2013,Fongetal2014,Barrotetal2016,Barberetal2023}. Consistent with these more recent findings, BJZZ empirically demonstrate that retail investors' order flows can predict future returns using U.S. equity market data between January 2010 to December 2015. Specifically, BJZZ write that ``\textit{(...) retail investors are slightly contrarian at a weekly horizon, and that the cross-section of weekly marketable retail order imbalances predicts the cross-section of returns over the next several weeks}'' \citep[p.2251]{Boehmeretal2021}.

To conduct their analysis, BJZZ develop an algorithm for identifying and signing retail trades with the NYSE Trade and Quote (TAQ) datasets. This method builds on the observation that retail trades are frequently executed off-exchange---by a wholesaler or through internalization---and often receive subpenny price improvements.
This approach offers a better alternative to previous methods that relied on trade size as a differentiator \citep[\eg,][]{Leeetal2000, Bhattacharyaetal2007, Campbelletal2009} or private brokerage data \citep[\eg,][]{BarberOdean2008, KelleyTetlock2013} and has quickly gained popularity in the literature.\footnote{According to Google Scholar, it has been cited by 424 articles as of mid-February 2024. See, for example, \citet{Blankespooretal2019, Busheeetal2020, Bonsalletal2020, Guest2021, Farrelletal2022, Doronetal2022, Barberetal2023, Bradleyetal2022}.}

Despite its popularity, the BJZZ approach faces criticism. \citet{Battalioetal2023} and \citet{Barberetal2023_algo} independently assess its accuracy. Based on proprietary data on retail and institutional trades from multiple sources, \citet{Battalioetal2023} identify both Type I (identifying non-retail trades as retail) and Type II errors (failure to correctly identify retail trades), concluding that the BJZZ's algorithm ``\textit{(...) identifies less than one-third of trades known to be retail and frequently could include known institutional trades as retail}'' \citep[][p.3]{Battalioetal2023}. \citet{Barberetal2023_algo} find that the BJZZ approach accurately identifies only 35\% of trades while incorrectly signing 28\% of those identified, based on their execution of 85,000 trades across six retail brokerage accounts between December 2021 and June 2022. In addition, they suggest an alternative method to identify and sign retail trades based on the \citet{LeeReady1991} quote midpoint (QMP) method. They note in the abstract of their paper that the QMP method ``\textit{(...) does not affect identification rates but reduces the signing error rates to 5\%.}'' 

The BJZZ and QMP methods share a similar procedure for identifying retail trades. Specifically, the identification process consists of (i) filtering for off-exchange transactions reported to a Financial Regulatory Authority (FINRA) Trade Reporting Facility (TRF)---these transactions are easily discernible in the TAQ datasets with the exchange code “D”; and (ii) isolating transaction prices that exhibit subpenny improvements, that is, those with a non-zero fraction of a penny, or third decimal. The approaches diverge in signing the trades. BJZZ classify a buy (sell) as any off-exchange transaction with a fractional cent between 0.6 and 1.0, exclusive (0.0 and 0.4, exclusive) and exclude transactions with a fractional cent between 0.4 and 0.6, inclusive. The QMP approach, on the other hand, signs a trade as a buy (sell) if the transaction price is greater (less) than the midquote price but does not sign trades whose price falls within 40\% and 60\% of the National Best Bid or Offer (NBBO). In other words, the QMP method considers spread dynamics, potentially leading to different trade classifications based on a stock’s spread size, contrary to the BJZZ method, which implicitly assumes a spread of exactly one penny. We refer to \citet{Boehmeretal2021} and \citet{Barberetal2023_algo} for more precision on each approach.

Our paper has two main objectives. First, we analyze whether BJZZ's original findings based on the 2010-2015 period continue to hold in the more recent 2016-2021 period, akin to an extension study. Second, we analyze the effect of using the alternative QMP method to identify and sign retail trades on their original conclusions in the original 2010-2015 and the more recent 2016-2021 periods. We should note that \citet{Barberetal2023_algo} provide evidence that the QMP method signs retail trades more accurately than the BJZZ approach and examine how this affects the estimation of retail order imbalances (see Section III of the appendix of their paper). However, a critical aspect that remains to be explored is whether the main conclusions of BJZZ on the predictive power of retail order imbalances for returns continue to hold when one uses the QMP method.

To address these objectives methodically, we begin by demonstrating that BJZZ's primary empirical results, as presented in their first eight tables (Tables I to VIII), can be replicated with high precision using their provided code.\footnote{We exclude Tables IX and X due to data availability reasons.} This replication serves as the foundation for our comparative analyses.

For the first objective---the extension study---we reproduce the first eight tables of BJZZ using data from the more recent period between 2016 and 2021. Our main results can be summarized as follows: In the recent 2016-2021 period compared to 2010-2015: (i) the empirical evidence for their findings that the main determinant of retail order imbalance (ROI) is its first lag is statistically much weaker; (ii) the original findings that past ROIs can predict next week returns are also statistically much weaker; (iii) the predictability patterns of large-cap and high-price stocks disappear, while those of small-cap and low-price stocks seriously weaken; (iv) ROI's ability to predict returns is confined mostly to four weeks instead   of the original six to eight weeks; (v) long-short strategies based on ROI are no longer profitable across all stocks and significantly less profitable for small stocks; (vi) the evidence supporting the notion that ROI's predictive power for returns is primarily due to the persistence of ROI weakens, albeit remaining significant; (vii) the lack of supporting evidence for the liquidity provision hypothesis persists and continues to conflict with the findings of \citet{KanielSaarTitman2008}. Overall, our results indicate either a substantial weakening or a disappearance of most of BJZZ's main findings in the recent 2016-2021 period.

For the second objective---analyzing the impact of the QMP method---we compute all retail-trade quantities based on the QMP instead of the BJZZ approach for both the original 2010-2015 and the recent 2016-2021 periods. We then reproduce BJZZ's original results and compare the two methods for each period. In the original period, most of BJZZ's empirical results continue to hold when using the QMP method. However, the QMP approach tends to provide stronger empirical support in the most recent period. This suggests that the fundamental differences between the two methods exert a more significant influence in recent times. This aligns with the correlations outlined in Table~\ref{tab:BJZZTAB1} of Section 3.1, indicating that ROIs based on share volume or number of trades using BJZZ and QMP approaches are highly correlated in the original period (68\% and 71\%, respectively), but substantially less correlated in the recent period (44\% and 53\%, respectively).

Based on the outcomes for two main objectives, we can draw a further comparison by contrasting the periods using the QMP method instead of BJZZ's. As anticipated, the empirical support for BJZZ's original conclusions also weakens in the recent 2016-2021 period when employing the QMP approach, albeit to a lesser degree.

Our paper makes at least three important contributions to the literature on retail investors. First, we demonstrate that BJZZ's original findings can be replicated with high precision. Second, we reveal that most of BJZZ's main findings either weaken significantly or disappear entirely in the recent 2016-2021 period. Third, we show that while the QMP method does not significantly alter BJZZ's main conclusions in the original 2010-2015 period, more pronounced differences emerge in the more recent 2016-2021 period.

The rest of the paper is organized as follows. Section 2 presents the data and the methodology we follow to construct our samples. Section 3 presents the results of our comparisons for each of BJZZ's first eight tables. Section 4 concludes. The code for reproducing our results will be made available on GitHub soon.

\section{Data and Methodology}

\noindent
We start by constructing two samples spanning from January 1, 2010, to December 31, 2021. In the first sample, we identify and sign retail trades following the BJZZ approach, utilizing the replication code provided by the authors.\footnote{BJZZ original replication code is accessible at \url{https://onlinelibrary.wiley.com/doi/abs/10.1111/jofi.13033}.} In the second sample, we implement the QMP approach based on our own code to identify and sign retail trades. We apply the data filters as specified by BJZZ to define the universe of stocks in both samples. Specifically, we retain only common stocks (CRSP's share codes 10 or 11) listed on the NYSE, NYSE MKT (formerly Amex), and NASDAQ, with a price of at least \$1 at the previous month-end. Following \citet{Barberetal2023}, we exclude stocks affected by the Tick Size Pilot program between October 2016 and October 2018.\footnote{Specifically, stocks from the test groups G2 and G3 are dropped. We identified these stocks using the \texttt{TICK\_PILOT\_INDICATOR} flag available in the TAQ datasets. See https://www.finra.org/rules-guidance/key-topics/tick-size-pilot-program for details.} 
We construct all variables necessary for reproducing Tables I-VIII in BJZZ. Non-retail-trade variables---stock return, market capitalization, turnover, book-to-market ratio, and volatility---are common to both samples, while retail-trade variables are sample-specific. Table~\ref{tab:VARDES} lists all variables and their acronyms used in our paper. 

\insertfloat{Table~\ref{tab:VARDES}}

Having constructed these two samples based on BJZZ and QMP approaches, we split each sample into two six-year periods: January 1, 2010, to December 31, 2015, representing the original BJZZ period, and January 1, 2016, to December 31, 2021, representing the recent period. The resulting four samples---referred to as \textit{(a) BJZZ 2010-2015}, \textit{(b) QMP 2010-2015}, \textit{(c) BJZZ 2016-2021}, and \textit{(d) QMP 2016-2021} for brevity---serve as the basis for our comparisons. Specifically, this framework enables four pairwise comparisons, two related to our first objective, examining the impact of the sample period (Panel (a) vs. (c) and Panel (b) vs. (d) in the various tables), and two related to our second objective, evaluating the effect of applying the QMP approach instead of the BJZZ approach on BJZZ's original conclusions (Panel (a) vs. (b) and Panel (c) vs. (d) in the various tables).

Before comparing periods and methods, we meticulously ensure the successful replication of BJZZ's original findings. Results of this replication exercise for each of the eight tables in BJZZ are reported in the online appendix, Tables~\ref{tab:BJZZTAB1_REP}-\ref{tab:BJZZTAB8_REP}. Overall, our results demonstrate that we can accurately replicate BJZZ's original results. In all subsequent tables, Panel \textit{(a) BJZZ 2010-2015} corresponds to these replication results.

\section{Results}

\noindent
This section presents our main findings, addressing our two objectives in tandem. Specifically, for each of the first eight tables of BJZZ, we discuss the results of our extension study---drawing distinctions between conclusions of the 2016-2021 and 2010-2015 periods---and the results from our investigations of the potential implications of employing the QMP  instead of the BJZZ approach. Each table comprises four panels corresponding to the four samples mentioned earlier. To ease the comparison with BJZZ's study, each table's number in this section aligns with the corresponding table's number in BJZZ. Moreover, we report only selected results from most tables to save space. The complete set of results is available in the online appendix; see Tables~\ref{tab:BJZZTAB2_COMPLETE}-\ref{tab:BJZZTAB8_COMPLETE}.

\subsection{Summary Statistics}

\noindent
Table~\ref{tab:BJZZTAB1} reports the summary statistics for our four samples. Comparing the recent and original periods (Panel (a) vs. (c) and Panel (b) vs. (d)) reveals an important increase in retail investor activity, as evidenced by both the daily average number of shares bought and sold (\textsf{Mrbvol} and \textsf{Mrsvol}) and the daily number of buy and sell transactions (\textsf{Mrbtrd} and \textsf{Mrstrd}). Furthermore, the daily mean and median of order imbalances (\textsf{Mroibvol} and \textsf{Mroibtrd}) are noticeably closer to zero in the recent period (\eg, \textsf{Mroibvol} of -0.018 in Panel (c) vs. -0.036 in (a)), suggesting that the heightened activity is predominantly driven by increased buying. When comparing the two approaches (Panel (a) vs. (b) and Panel (c) vs. (d)), the daily means and medians of \textsf{Mrbvol}, \textsf{Mrsvol}, \textsf{Mrbtrd}, and \textsf{Mrstrd} based on QMP are consistently higher than those based on BJZZ in both periods. This indicates that QMP captures a higher average trading activity than BJZZ. Additionally, QMP yields more negative order imbalances on average (\eg, \textsf{Mroibvol} of -0.036 in Panel (d) vs. -0.018 in (c)), suggesting that QMP might be better at identifying sell transactions than BJZZ. We compute the correlations between the QMP-based and BJZZ-based quantities to scrutinize the differences between the two approaches further. Interestingly, the correlations between the order imbalance measures decrease significantly in the recent period, from 0.68 (0.71) in 2010-2015 to 0.44 (0.53) in 2016-2021 for \textsf{Mroibvol} (\textsf{Mroibtrd}). This finding might be an early indication that the potential divergence in results based on the BJZZ and QMP approaches could intensify in more recent years.

\insertfloat{Table~\ref{tab:BJZZTAB1}}

\subsection{Determinants of Marketable Retail Order Imbalances}

\noindent
Table~\ref{tab:BJZZTAB2} reports the results on the determinants of marketable retail order imbalances (ROI), analogous to Table II in BJZZ. BJZZ investigate the relationship between retail investors' marketable order flow and past order flow, as well as past returns.
They regress the current-week order imbalance for a given stock $i$ ($\textsf{Mroib}(i,w)$) on the previous-week order imbalance ($\textsf{Mroib}(i,w-1)$), previous-week returns ($\textsf{Ret}(i,w-1)$), and various control variables ($\textsf{CTRL$(i,w-1)$}$) including previous-month returns ($\textsf{Ret}(i,m-1)$), previous-six-months returns prior to the last month ($\textsf{Ret}(i,m-7,m-2)$), last-month-end turnover ($\textsf{Lmto$(i,m-1)$}$), last-month volatility ($\textsf{Lvol$(i,m-1)$}$), last-month-end size ($\textsf{Size$(i,m-1)$}$), and last-month-end book-to-market ($\textsf{Lbm$(i,m-1)$}$).\footnote{Note that, while the variables \textsf{Mroib} and \textsf{Ret} are originally measured at the daily level, the analyses of BJZZ ``\textit{focus on weekly horizons to reduce the impact of microstructure noise on [their] results}'' \citep[p.2262]{Boehmeretal2021}. Also, in what follows, all discussions pertain to order imbalances based on share volume (\textsf{Mroibvol} for \textsf{Mroib}) and bid-ask average returns (\textsf{Ret}). Using order imbalances based on the number of trades (\textsf{Mroibtrd}) and/or CRSP closing price returns does not fundamentally change the interpretation; see the online appendix.} 
They employ the \citetalias{FamaMcBeth1973} approach to analyze this relation. Specifically, in the first stage, for each day, they estimate the following cross-section regression: 
\begin{align}\label{eq:Reg_DETERMINANTS} 
\begin{split}
\textsf{Mroib}(i,w)
& = b_0(w) + b_1(w) \textsf{Mroib}(i,w-1) + b_2(w) \textsf{Ret}(i,w-1) \\
& \quad + b_3(w)' \textsf{CTRL}(i,w-1) + u_1(i,w) \,. \\
\end{split}
\end{align}
In the second-stage, they conduct statistical inference using the time-series of the coefficients, $\{b_0(w), b_1(w), b_2(w),b_3(w)'\}$ and \citet{NW1987} standard errors with six lags.\footnote{In all tables, we compute standard errors following the method and lag specifications outlined in BJZZ. Specifically, for Table II (Tables III, IV, V, and VII), BJZZ use Newey-West standard errors with six (five) lags. We also consider a lag length of 10 and our results are very similar to those presented.}

BJZZ's original results suggest that the primary determinant of weekly ROI is its first lag. Our results in Panel (c) show that this conclusion still holds in 2016-2021, although its economic and statistical significance weaken. For instance, the coefficient estimate and the $t$-stat of the first lag, $\textsf{Mroibvol}(i,w-1)$, are 50\% (0.0983 vs. 0.1982) and 20\% (57.30 vs. 71.81) lower in 2016-2021 period, respectively. When we compare periods using the QMP method instead (Panel (d) vs. (b)), this weakening effect is also visible, although to a lesser extent.

Employing the QMP method does not materially alter BJZZ's original finding, as the $\textsf{Mroibvol}(i,w-1)$ estimate and $t$-stat are only slightly higher than those based on the BJZZ method in 2010-2015 (0.2360 and 84.01 in Panel (b) vs. 0.1982 and 71.81 in (a)). In the more recent period, however, the divergences increase, with $\textsf{Mroibvol}(i,w-1)$ estimate and $t$-stat much higher for QMP (0.1729 and 80.38, Panel (d)) than for BJZZ (0.0983 and 57.30, Panel (c)). Therefore, the QMP-based results provide similar evidence for BJZZ's finding in the original period but stronger evidence in 2016-2021.

\insertfloat{Table~\ref{tab:BJZZTAB2}}

\subsection{Predicting Next-Week Returns Using Marketable Retail Order Imbalances}

\noindent
Table~\ref{tab:BJZZTAB3} presents results on the predictability of next-week returns using marketable retail order imbalances, analogous to Table III in BJZZ. To perform this analysis, BJZZ regress current-week returns ($\textsf{Ret}(i,w)$) on previous-week order imbalances ($\textsf{Mroib}(i,w-1)$). Regressions include the same controls as in \eqref{eq:Reg_DETERMINANTS}, with the addition of $\textsf{Ret}(i,w-1)$. Again, they estimate this regression using the \citetalias{FamaMcBeth1973} approach, where the first-stage cross-section regressions are given by:
\begin{align}\label{eq:Reg_PREDICT} 
\begin{split}
\textsf{Ret}(i,w) = c_0(w) + c_1(w)\textsf{Mroib}(i,w-1) + c_2(w)'\textsf{CTRL}(i,w-1) +u_2(i,w) \,. \\
\end{split}
\end{align}
In the second-stage, they conduct statistical inference using the time-series of the coefficients, $\{c_0(w), c_1(w),c_2(w)'\}$ and \citet{NW1987} standard errors with five lags. For more details, see \citet[pp.2266-2267]{Boehmeretal2021}.

BJZZ's original results indicate that past-week ROIs can predict future returns in the same direction, that is, the coefficient $\hat{c_1}$ is significantly positive. Our results based on the BJZZ method (Panel (c)) in the recent period reveal a considerably weaker predictive power compared to the BJZZ's original findings in 2010-2015. Specifically, the coefficient estimate of $\textsf{Mroibvol}(i,w-1)$ and its $t$-stat are 34\% (0.0006 vs. 0.0009) and 48\% (7.95 vs. 15.14) lower, respectively, and the corresponding economic magnitude decreases from 11.16 basis points per week (or $0.1116\% \times 52 = 5.8\%$ per year) to 6.02 basis points per week (3.1\% per year). This finding holds true when using QMP but to a lesser extent with the economic magnitude decreasing only from 6.3\% to 5.6\% per year (Panel (d) vs. (b)). 

Comparing BJZZ and QMP methods in 2010-2015, predictability holds with the same order of economic magnitude for both methods (Panel (b) vs. (a)). In 2016-2021, however, QMP tends to reinforce the evidence for predictability (Panel (d) vs. (c)). For example, the coefficient estimate on $\textsf{Mroibvol}(i,w-1)$ and its $t$-stat are 63\% and 28\% higher, respectively, corresponding to an economic magnitude increasing from 3.1\% to 5.6\% per year. We should note that we follow this example and base our discussions on the order imbalances based on share volume (\textsf{Mroibvol}) for the rest of the paper. Using order imbalances based on the number of trades (\textsf{Mroibtrd}) does not fundamentally change our main conclusions. 

\insertfloat{Table~\ref{tab:BJZZTAB3}}

\subsection{Marketable Retail Return Predictability Within Subgroups}

\noindent
Table~\ref{tab:BJZZTAB4} reports results about marketable retail return predictability within subgroups, analogous to Table IV in BJZZ. In this analysis, BJZZ explore questions such as (p.2267): ``\textit{(...) is the predictive power of marketable retail order imbalances restricted to a particular type of firm?}'' or ``\textit{(...) do informed retail investors have preferences for particular types of firms?}'' To address them, they construct subgroups based on three characteristics---market capitalization, share price, and turnover, all calculated at the previous month-end---and estimate \eqref{eq:Reg_PREDICT} within each characteristic group. For more details, see \citet[p.2267]{Boehmeretal2021}.

BJZZ's original results indicate that predictability exists for all market-cap, share-price, and turnover groups. Furthermore, within these groups, they observe stronger predictability for small-cap, low-price, and low-turnover stocks. Reproducing the results for the recent period with the BJZZ approach (Panel (c) vs. (a)) reveals that the original conclusions tend to weaken or disappear in 2016-2021. Indeed, the economic magnitude associated with the predictability of small-cap (low-price) stocks decreases from 10.7\% to 5.1\% (10.7\% to 6.9\%) per year. Additionally, the statistically significant (at the 1\% significance level) predictive power for big-cap and high-price stocks that existed in 2010-2015 completely disappears in 2016-2021. When we compare periods using QMP instead (Panel (d) vs. (b)), the predictive power still weakens or disappears in the recent period, but to a lesser extent. For example, the economic magnitude associated with the predictability of returns on small-cap stocks decreases only from 11.6\% to 10.2\%, and the predictability of returns on high-price stocks continues to hold.

Contrasting results between methods show that both yield similar results in 2010-2015 (Panel (b) vs. (a)). In 2016-2021, however, important differences arise with the QMP approach suggesting stronger predictability for most subgroups (Panel (d) vs. (c)). For instance, the economic magnitude associated with the predictability of returns on small-cap stocks is approximately twice as large with QMP (10.2\% vs. 5.1\%).

\insertfloat{Table~\ref{tab:BJZZTAB4}}

\subsection{Predicting Returns $k$-Weeks Ahead}

\noindent
Table~\ref{tab:BJZZTAB5} reports results on $k$-weeks ahead predictions, analogous to Table V in BJZZ. Specifically, they analyze the predictive power of marketable retail order imbalances at horizons longer than one week, aiming to discern whether the predictive power is transient or persistent. They state, ``\textit{(...) if the predictive power quickly reverses, the retail investors may be capturing price reversals; if the predictive power continues over time and then vanishes beyond some horizon, the retail investors may be informed about information related to firm fundamentals''} \citep[p.2270]{Boehmeretal2021}. They address this question by making slight adjustments to \eqref{eq:Reg_PREDICT}, allowing for horizons of $k>1$ weeks. The first stage of their \citetalias{FamaMcBeth1973} estimation becomes:

\begin{align}\label{eq:Reg_PREDICTHORIZON} 
\begin{split}
\textsf{Ret}(i,w+k) = c_0(w) + c_1(w) \textsf{Mroib}(i,w) + c_2(w)'\textsf{CTRL}(i,w) +u_3(i,w+k), \\
\end{split}
\end{align}
where they allow $k$ to vary from one to 12 weeks, and $\textsf{Ret}(i,w+k)$ represents the \emph{one-week period} return $k$ week ahead, rather than a cumulative return over $k$ weeks. For more details, see \citet[pp.2270-2271]{Boehmeretal2021}. 

BJZZ's original results indicate that retail order imbalances can predict future returns up to six to eight weeks ahead. In addition, they observe that the predictive power generally decreases monotonically with the horizon. BJZZ's original conclusions tend to weaken or disappear in 2016-2021 (Panel (c) vs. (a)). In 2010-2015, for instance, ROI can predict returns up to eight weeks ahead (\eg, eight-week \textsf{Mroibvol} coefficient of 0.0002 with a $t$-stat of 3.96). In the recent period, however, the predictive significance starts to weaken at four weeks and beyond. Indeed, the four- and six-week \textsf{Mroibvol} coefficients of 0.0002 and 0.0002 are significant at the 5\% level only, and the eight weeks coefficient of 0.0001 is significant at the 10\% level only. When comparing periods using the QMP method (Panel (d) vs. (b)), this interpretation holds true but to a lesser extent, as \textsf{Mroibvol}'s predictive power loses statistical significance rather after six weeks than four weeks.

Regarding the comparison between methods (Panel (b) vs. (a) and Panel (d) vs. (c)), both lead to similar conclusions and economic magnitudes in 2010-2015, but notable differences arise in 2016-2021, with stronger and more significant predictive coefficients for all horizons when using QMP. For example, in 2016-2021, the coefficient on \textsf{Mroibvol} at the two-week horizon and its $t$-stat are 0.0004 and 4.19 with QMP compared to 0.0003 and 3.46 with BJZZ. At the four-week horizon, they are respectively 0.0004 and 3.91 with QMP compared to 0.0002 and 2.02 with BJZZ.

\insertfloat{Table~\ref{tab:BJZZTAB5}}

\subsection{Long-Short Strategy Returns Based on Marketable Retail Order Imbalances}

\noindent
Table~\ref{tab:BJZZTAB6} reports results about long-short strategy returns based on marketable retail order imbalances, analogous to Table VI in BJZZ.
Specifically, BJZZ analyze whether marketable retail order imbalances can be used as a signal to form a profitable trading strategy. Their insight is that ``\textit{(...) if retail investors on average can select the right stocks to buy and sell, then firms with higher or positive marketable retail order imbalance should outperform firms with lower or negative order imbalance''} \citep[p.2271]{Boehmeretal2021}. To address this question, they form quintile portfolios based on the average order imbalance over the previous week and construct long-short portfolios where the stocks in the highest order imbalance quintile are bought and the stocks in the lowest order imbalance quintile are shorted. The performance of the long-short portfolios is assessed in terms of raw and risk-adjusted returns (\ie, alpha) against the \citet{FF1993} three-factor model, and for horizons up to 12 weeks. The returns are value-weighted using previous month-end market cap.\footnote{Note that they further precise: ``\textit{Notice that this exercise uses marketable retail order imbalance measures merely as a signal to predict future stock returns, and thus, it provides no information on whether retail investors with marketable orders profit from their own trades. We ignore trade frictions and transaction costs here, and thus, the results do not have implications for whether outsiders can profit from these signals}'' \citep[p.2271]{Boehmeretal2021}.}

BJZZ's original results indicate that such a long-short strategy generates statistically positive alphas at horizons from one to 12 weeks, and that results are more pronounced with a universe of small-cap stocks only. In 2016-2021, this strategy ceases to be profitable when we consider all stocks available in that sample period. Indeed, results in the recent period (Panels (c) and (d)) show that the alphas based on the universe of all stocks are no longer statistically significant in the recent period, regardless of the horizon or the method considered. For a strategy on small-cap stocks only, alphas remain statistically positive but experience a notable decline. Specifically, comparing BJZZ samples (Panel (c) vs. (a)) reveals that at all horizons, small-sample alphas and their $t$-stat are much lower compared to the 2010-2015 period. For example, for one- and two-week horizons, 2016-2021 alphas are more than 65\% lower (0.143\% vs. 0.437\% and 0.177\% vs. 0.613\%, respectively). Moreover, at horizons of eight weeks and beyond, the 2016-2021 small-cap alphas are no longer significant. Based on the QMP method (Panel (d) vs. (b)), the small-sample alphas also exhibit a significant decrease, albeit less pronounced, and lose statistical significance at one more horizon step (\ie, 10 weeks).

Turning to the comparison between methods, in general, both lead to similar conclusions and economic magnitudes in both periods (Panel (b) vs. (a) and Panel (d) vs. (c))---with some notable differences in the recent period such as significantly stronger one-week small cap alpha for the QMP method (0.295\% vs. 0.143\%).

\insertfloat{Table~\ref{tab:BJZZTAB6}}

\subsection{Predictability Decomposition}

\noindent
Table~\ref{tab:BJZZTAB7} presents results regarding predictability decomposition, analogous to Table VII in BJZZ. In this analysis, BJZZ explore three alternative hypotheses that could elucidate why marketable retail order imbalance can predict future returns. The first hypothesis hinges on the persistence of order flows \citep[see \eg,][]{Chordiaetal2004}. The second hypothesis relies on the contrarian trading behavior exhibited by retail investors \citep[see \eg,][]{KanielSaarTitman2008}. The third hypothesis posits that retail investors may accurately predict the direction of future returns because they possess valuable information about the firm \citep[see \eg,][]{KelleyTetlock2013}. To test these hypotheses, BJZZ adopt a two-stage decomposition. First, they decompose their retail marketable order imbalance variable in three parts: $\textsf{Mroib}(i,w) = \widehat{\textsf{Mroib}^{persistence}_{i,w}} + \widehat{\textsf{Mroib}^{contrarian}_{i,w}} + \widehat{\textsf{Mroib}^{other}_{i,w}}$. Then, they estimate \eqref{eq:Reg_PREDICT} where $\textsf{Mroib}(i,w-1)$ is replaced by these three components:
\begin{align}\label{eq:Reg_PREDICTDECOMPOSITION} 
\begin{split}
\textsf{Ret}(i,w) 
& = e_0(w) + e_1(w) \widehat{\textsf{Mroib}^{persistence}_{i,w-1}} + e_2(w) \widehat{\textsf{Mroib}^{contrarian}_{i,w-1}} \\
& \quad + e_3(w) \widehat{\textsf{Mroib}^{other}_{i,w-1}} + e_4(w)' \textsf{CTRL}(i,w-1) +u_5(i,w) \,. \\
\end{split}
\end{align}
For more details, see \citet[pp.2274-2277]{Boehmeretal2021}.

BJZZ's original results indicate that most of the predictability primarily comes from the persistence (PERS) and residual (OTHER) components of retail order imbalance---the latter aligning with the third hypothesis described above, that is, marketable retail investor trading contains valuable information about future stock price movements. They also show that the contrarian trading pattern component (CONT) lacks statistical significance. The persistence and residual components, though still significant, seriously weaken in the recent period. Specifically, based on the BJZZ samples (Panel (c) vs. (a)), the significance of PERS drops from 1 to 10\%, and its economic magnitude decreases from 0.0692\% (3.60\% per year) to 0.0241\% (1.25\% per year); and the OTHER coefficient decreases from 0.0008 ($t$-stat of 13.02) to 0.0006 ($t$-stat of 7.68). If we rather consider the QMP samples (Panel (d) vs. (b)), we observe a similar trend, albeit less pronounced. Finally, regardless of the method used, the contrarian component remain statistically insignificant in 2016-2021. 

When comparing methods (Panel (b) vs. (a) and Panel (d) vs. (c)), results are similar in 2010-2015. However, in 2016-2021, the persistence and residual components show greater statistical significance and larger economic magnitudes when we rely on QMP. For instance, PERS and OTHER based on \textsf{Mroibvol} in Panel (d) have economic magnitudes of 0.0534\% and 0.0941\%, respectively, compared to 0.0241\% and 0.0580\% in Panel (c).

\insertfloat{Table~\ref{tab:BJZZTAB7}}

\subsection{Marketable Retail Order Imbalance and Contemporaneous Returns}

\noindent
Table~\ref{tab:BJZZTAB8} reports results about marketable retail order imbalance and contemporaneous returns, analogous to Table VIII in BJZZ. In this analysis, BJZZ explore the liquidity provision hypothesis, relying on the work of \citet[KST]{KanielSaarTitman2008} who argue that ``\textit{(...) retail investors' contrarian trading provides liquidity to the market and leads to the positive predictive power of past marketable retail order imbalances for future stock returns}'' \citep[p.2281]{Boehmeretal2021}. Specifically, in their Table III, KST examine the past, contemporaneous, and future returns of intense buy and sell portfolios of retail investors, where these portfolios are constructed based on previous week's net individual trading (NIT)---the KST equivalent measure of retail trading flows. KST's findings are threefold: (i) they observe typical contrarian trading behavior by retail investors;\footnote{That is, ``\textit{(...) the stocks that retail investors sell during the portfolio construction week (week 0)—the intense selling group—experience significantly positive excess returns before week 0, while the stocks that retail investors buy during week 0—the intense buying group—experience negative excess returns.}'' \citep[][p.2282]{Boehmeretal2021}} (ii) they show that retail trading can predict returns in the correct direction;\footnote{That is, ``\textit{(...) after retail investors buy or sell, the stocks that retail investors sell during week 0 (the intense selling group) experience negative excess returns, while the stocks that retail investors buy during week 0 (the intense buying group) experience positive excess returns.}'' \citep[p.2282]{Boehmeretal2021}} and (iii) they find that ``\textit{(...) the contemporaneous excess return is significantly positive for stocks that retail investors sell and negative for stocks that they buy}'' \citep[p.2282]{Boehmeretal2021}, which is in favor of the liquidity provision hypothesis.

To test for the liquidity provision hypothesis, BJZZ replicate Table III of KST, where they construct portfolios using their own retail order imbalance measures, \textsf{Mroibvol} and \textsf{Mroibtrd}. BJZZ's findings validate the first two observations of KST: the contrarian trading patterns of retail investors and the predictive power of order imbalance on future returns. However, BJZZ's results diverge from KST's third assertion concerning the liquidity provision hypothesis, with no supporting evidence found in BJZZ's findings, contrary to KST's. In the recent period, we observe notable changes in contrarian behaviors, primarily on the buying side. Retail investors still tend to buy stocks with negative returns but do not consistently sell stocks with positive returns. Indeed, results for both the BJZZ samples (Panel (c) vs. (a)) and QMP samples (Panel (d) vs. (b)) show low or insignificant $t$-stats for Intense Selling portfolios for $k<0$ (see \eg, coefficients corresponding to $k=-15,-10,-5$ in Panel (c)). Regarding predictive power, the 2016-2021 results largely echo those of 2010-2015, with a noteworthy decline in significance specific to the buying side, dropping from 1 to 5\% (see $k>0$ ``Intense Buying'' estimates in Panel (c) vs. (a)). Finally, aligning with BJZZ's 2010-2015 results, we do not find evidence supporting the liquidity provision hypothesis in the recent period, as $k=0$ estimates are either zero or align with the trade direction. 

When comparing methods, results align in 2010-2015, and slight variations emerge in 2016-2021. For instance, QMP tends to provide more supporting evidence for contrarian behavior on the selling side compared to the BJZZ approach (\eg, $k=-10$ and $k=-5$ estimates are significant at the 5\% or 1\% levels in Panel (d), while they are not statistically significant in Panel (c)).

\insertfloat{Table~\ref{tab:BJZZTAB8}}

\section{Conclusion}

\noindent
In this study, we offer new insights into the relationship between retail investors' trading activity and future stock returns, a subject of considerable interest in finance literature. Our research centers on revisiting the findings of \citet[BJZZ]{Boehmeretal2021}. This paper, cited by 424 studies as of mid-February 2024, has garnered substantial attention in the finance literature due to its compelling findings and methodological innovation, introducing a novel algorithm for identifying retail trades within the NYSE Trade and Quote (TAQ) datasets.

Our study focuses on two principal objectives. First, to appraise the persistence of BJZZ's original 2010-2015 findings regarding the predictive power of retail order imbalances (ROI) on future stock returns into the more recent 2016-2021 period. Second, to evaluate the impact of an alternative method to identify and sign retail trades---specifically the \citet{LeeReady1991} quote midpoint (QMP) method recommended in \citet{Barberetal2023_algo}---on statistical inferences.

To achieve these goals, we first replicate BJZZ's original findings with high precision using their provided code. Then, we extend the analysis to 2016-2021 and construct additional samples where retail-trades quantities are computed using the QMP instead of the BJZZ method.

Regarding the first objective, notable differences emerge in 2016-2021, with a marked reduction in the strength of several key findings regarding the predictive ability of retail order imbalance on future returns. Notably, the predictability for large-cap and high-price stocks vanishes, and that for small-cap and low-price stocks seriously weakens. Additionally, the profitability of long-short strategies based on past ROI disappears in a universe of all stocks and substantially decreases in a universe of small-cap stocks only.
Regarding the second objective---contrasting results when employing the QMP instead of the BJZZ approach---we first notice a significant drop in the correlation between BJZZ-based and QMP-based order imbalances series in recent years, from 68\% in 2010-2015 to 44\% in 2016-2021. Consistent with this observation, our results also indicate that while both methods yield similar conclusions in the original 2010-2015 period, divergences increase in the recent 2016-2021 period, with the QMP method lending stronger support to BJZZ's original findings.

Our study makes three contributions to the literature on retail investors. First, we successfully replicate BJZZ's original findings with high precision. Second, we highlight that changing market dynamics in the recent 2016-2021 period significantly impact the predictive power of retail investors' trading patterns. Lastly, we find that using the QMP method instead of the BJZZ method in their original period does not significantly alter BJZZ's original results. Yet, in recent years, the differences between the methods have a more noticeable effect on predictive regression outcomes.

In conclusion, our study confirms the validity of BJZZ's methodology and findings in their original context while raising essential questions about the temporal stability of these findings. Evolving market conditions appear to have diluted the predictive power of retail investors' trading decisions. Additionally, the use of the alternative QMP method materially impacts BJZZ's initial findings, particularly in recent years. Overall, our findings underscore the necessity for a continuous reassessment of methodologies and conclusions in the rapidly evolving landscape of financial markets.

\newpage
\bibliographystyle{elsarticle-harv}

\setcounter{table}{0}
\renewcommand*{\thetable}{\Alph{table}} 
\renewcommand{\theHtable}{Supplement.\thetable}

\newpage
\begin{table}[h!]
\caption{\textbf{Description of Variables}\\
\textbf{Description}: This table lists all the variables used in our analyses. Non-retail trade variables are common to BJZZ and QMP samples, while retail trade variables are sample-specific.} 
\label{tab:VARDES}
\centering
\scalebox{0.93}{
    \centering
    \begin{tabular}{ll}
    \toprule
    Variable & Description \\ \hline
        \multicolumn{2}{l}{\textit{Non-Retail-Trades Variables Common to BJZZ and QMP Samples}} \\
        \textsf{Ret} & Bid-ask average return \\ 
        \textsf{Lmto} & Last-month-end turnover \\ 
        \textsf{Size} & Last-month-end logarithm of market value  \\ 
        \textsf{Lbm} & Last-month-end logarithm of book-to-market \\ 
        \textsf{Lvol} & Last-month volatility of daily returns \\
        \\
        \multicolumn{2}{l}{\textit{Retail-Trades Variables Specific to BJZZ or QMP Samples}} \\
        \textsf{Mrbvol} & Marketable retail buy volume based on shares traded \\ 
        \textsf{Mrsvol} & Marketable retail sell volume based on shares traded \\ 
        \textsf{Mrbtrd} & Marketable retail number of buy trades \\ 
        \textsf{Mrstrd} & Marketable retail number of sell trades \\ 
        \textsf{Mroibvol} & Marketable retail order imbalance based on shares traded \\ 
        \textsf{Mroibtrd} & Marketable retail order imbalance based on number of trades \\ 
    \bottomrule
    \end{tabular}}
\end{table}

\setcounter{table}{0}
\renewcommand*{\thetable}{\arabic{table}} 

\begin{landscape}
\begin{table}[!h]
\caption{\textbf{Summary Statistics} \\
\textbf{Description}: This table presents selected summary statistics analogous to Table I of BJZZ. To save space, we only report results based on round lots and odd lots. In Panel (a), we present statistics derived from our replication of their original sample. In Panels (b), (c), and (d), we report statistics derived from our samples in the recent period and those utilizing the QMP method.
\\ 
\textbf{Interpretation}: Retail investors' daily activity has increased in recent years, predominantly propelled by increased buying. The QMP approach appears to capture a higher activity than BJZZ. Additionally, QMP yields more negative order imbalances, suggesting it identifies more sell transactions than BJZZ. Most importantly, the correlations between the QMP- and BJZZ-based order imbalance measures significantly decrease in 2016-2021.}
\label{tab:BJZZTAB1}
\centering
\scalebox{0.93}{
    \centering
    \begin{tabular}{lccccccccccccc}
    \toprule
     & \multicolumn{6}{c}{\textit{(a) BJZZ 2010-2015}}  & \multicolumn{6}{c}{\textit{\textit{(b) QMP 2010-2015}}} & \textit{Correlation} \\
     \cmidrule(lr){2-7} \cmidrule(lr){8-13} \cmidrule(lr){14-14}
         & N & Mean & Std & Median & Q1 & Q3 & N & Mean & Std & Median & Q1 & Q3 \\ \cmidrule(lr){1-7} \cmidrule(lr){8-13}
        \textsf{Mrbvol} & 4,348,327 & 39,840 & 278,026 & 4,899 & 1,130 & 19,165 & 4,383,761 & 42,359 & 290,669 & 5,832 & 1,370 & 21,939 & 0.99 \\ 
        \textsf{Mrsvol} & 4,348,327 & 39,655 & 262,689 & 5,302 & 1,300 & 20,185 & 4,383,761 & 42,097 & 270,097 & 6,405 & 1,600 & 23,057 & 0.99 \\ 
        \textsf{Mrbtrd} & 4,348,327 & 99 & 386 & 21 & 5 & 72 & 4,383,761 & 112 & 406 & 26 & 6 & 87 & 1.00 \\ 
        \textsf{Mrstrd} & 4,348,327 & 97 & 330 & 22 & 6 & 74 & 4,383,761 & 110 & 350 & 28 & 7 & 90 & 0.99 \\ 
        \textsf{Mroibvol} & 4,348,327 & $-$0.036 & 0.470 & $-$0.025 & $-$0.304 & 0.224 & 4,383,761 & $-$0.045 & 0.460 & $-$0.032 & $-$0.304 & 0.208 & 0.68 \\ 
        \textsf{Mroibtrd} & 4,348,327 & $-$0.031 & 0.443 & $-$0.007 & $-$0.280 & 0.213 & 4,383,761 & $-$0.034 & 0.430 & $-$0.014 & $-$0.269 & 0.200 & 0.71 \\ 
        \midrule
     & \multicolumn{6}{c}{\textit{(c) BJZZ 2016-2021}}  & \multicolumn{6}{c}{\textit{\textit{(d) QMP 2016-2021}}} & \textit{Correlation} \\
     \cmidrule(lr){2-7} \cmidrule(lr){8-13} \cmidrule(lr){14-14}
         & N & Mean & Std & Median & Q1 & Q3 & N & Mean & Std & Median & Q1 & Q3 \\ \cmidrule(lr){1-7} \cmidrule(lr){8-13}
        \textsf{Mrbvol} & 3,965,568 & 55,607 & 435,949 & 5,667 & 1,390 & 21,394 & 3,968,258 & 57,459 & 438,907 & 6,557 & 1,642 & 23,597 & 1.00 \\ 
        \textsf{Mrsvol} & 3,965,568 & 54,750 & 416,327 & 5,928 & 1,480 & 22,144 & 3,968,258 & 56,211 & 414,076 & 7,032 & 1,875 & 24,535 & 1.00 \\ 
         \textsf{Mrbtrd} & 3,965,568 & 218 & 1573 & 37 & 11 & 115 & 3,968,258 & 239 & 1699 & 44 & 13 & 134 & 0.99 \\ 
        \textsf{Mrstrd} & 3,965,568 & 195 & 1256 & 38 & 11 & 114 & 3,968,258 & 205 & 1214 & 45 & 14 & 131 & 0.99 \\ 
         \textsf{Mroibvol} & 3,965,568 & $-$0.018 & 0.417 & $-$0.011 & $-$0.240 & 0.200 & 3,968,258 & $-$0.036 & 0.416 & $-$0.022 & $-$0.262 & 0.189 & 0.44 \\ 
        \textsf{Mroibtrd} & 3,965,568 & $-$0.005 & 0.326 & 0.000 & $-$0.154 & 0.149 & 3,968,258 & $-$0.007 & 0.336 & 0.000 & -0.172 & 0.171 & 0.53 \\         
    \bottomrule
    \end{tabular}}
\end{table}
\end{landscape}

\newgeometry{left=2cm,right=2cm,top=2cm,bottom=2cm}
\begin{table}[!h]
\caption{\textbf{Determinants of Marketable Retail Order Imbalances} \\
\textbf{Description}: This table displays results analogous to Table II of BJZZ. Specifically, BJZZ investigate the relationship between retail investors' marketable order flow and past order flow through the following \citetalias{FamaMcBeth1973} two-stage estimation: $\textsf{Mroib}(i,w) = b_0(w) + b_1(w) \textsf{Mroib}(i,w-1) + b_2(w) \textsf{Ret}(i,w-1) + b_3(w)' \textsf{CTRL}(i,w-1) + u_1(i,w)$. The (second-stage) standard errors' estimates are calculated using \citetalias{NW1987} with six lags. In Panel (a), we report our replication of their original findings. In Panels (b), (c), and (d), we revisit them using a more recent period and the alternative QMP method. To save space, we only report results based on bid-ask returns.
\\
\textbf{Interpretation*}: BJZZ's original finding (Panel (a)) is that the primary determinant of weekly ROI is its first lag. Comparing periods with the BJZZ method (Panel (c) vs. (a)) suggests that the evidence supporting BJZZ's original finding significantly weakens in the recent period. For example, the $\textsf{Mroibvol}(w-1)$ coefficient decreases from 0.1982 in 2010-2015 to 0.0983 in 2016-2021. Its significance also weakens, with $t$-stats of 71.81 and 57.30 in the original and recent period, respectively. When we compare periods using the QMP method instead (Panel (d) vs. (b)), this weakening effect is also visible, although to a lesser extent. Employing the QMP method does not alter materially BJZZ's original finding (Panel (b) vs. (a)). However, in the more recent period (Panel (d) vs. (c)), the divergences increase, with $\textsf{Mroibvol}(w-1)$ estimate and $t$-stat much higher for QMP than for BJZZ (0.1729 and 80.38 vs. 0.0983 and
57.30, respectively).}
\label{tab:BJZZTAB2}
\centering
\scalebox{0.9}{
    \centering
    \begin{tabular}{lrrrrrrrr}
    \toprule
     & \multicolumn{4}{c}{\textit{(a) BJZZ 2010-2015}}  & \multicolumn{4}{c}{\textit{(b) QMP 2010-2015}} \\
     \cmidrule(lr){2-5} \cmidrule(lr){6-9}
     & \multicolumn{2}{c}{\textsf{Mroibvol}} 
     & \multicolumn{2}{c}{\textsf{Mroibtrd}}
     & \multicolumn{2}{c}{\textsf{Mroibvol}} 
     & \multicolumn{2}{c}{\textsf{Mroibtrd}}\\ 
     \cmidrule(lr){2-3}\cmidrule(lr){4-5}\cmidrule(lr){6-7}\cmidrule(lr){8-9}
         & Coef & \textit{t}-stat & Coef & \textit{t}-stat & Coef & \textit{t}-stat & Coef & \textit{t}-stat \\
\midrule
        Intercept & $-$0.2833 & $-$22.23 & $-$0.2866 & $-$21.02 & $-$0.3593 & $-$22.43 & $-$0.2998 & $-$21.00 \\ 
        \textsf{Mroib}($w-1$) & 0.1982 & 71.81 & 0.2698 & 91.06 & 0.2360 & 84.01 & 0.2889 & 92.19 \\ 
        \textsf{Ret}($w-1$)  & $-$0.8302 & $-$35.91 & $-$0.7782 & $-$31.41 & $-$0.9157 & $-$38.83 & $-$0.8286 & $-$34.99 \\ 
        \textsf{Ret}($m-1$) & $-$0.1680 & $-$13.08 & $-$0.1214 & $-$8.91 & $-$0.1599 & $-$12.39 & $-$0.0730 & $-$5.74 \\ 
        \textsf{Ret}($m-7,m-2$) & $-$0.0252 & $-$5.20 & $-$0.0080 & $-$1.44 & $-$0.0163 & $-$3.25 & 0.0098 & 1.81 \\ 
        \textsf{Lmto} & 0.0007 & 11.73 & 0.0006 & 9.52 & 0.0009 & 14.15 & 0.0006 & 10.24 \\ 
        \textsf{Lvol} & 0.5684 & 6.43 & 0.3049 & 3.18 & 0.7436 & 7.84 & 0.1224 & 1.31 \\ 
        \textsf{Size} & 0.0151 & 10.89 & 0.0200 & 14.36 & 0.0220 & 13.72 & 0.0225 & 15.39 \\ 
        \textsf{Lbm} & $-$0.0211 & $-$16.95 & $-$0.0218 & $-$17.77 & $-$0.0264 & $-$21.48 & $-$0.0242 & $-$21.75 \\ 
        Adj.$R^2$ & 5.06\% & ~ & 8.66\% & ~ & 7.11\% & ~ & 9.90\% & ~ \\ 
        \midrule
        & \multicolumn{4}{c}{\textit{(c) BJZZ 2016-2021}}  & \multicolumn{4}{c}{\textit{(d) QMP 2016-2021}} \\
     \cmidrule(lr){2-5} \cmidrule(lr){6-9}
     & \multicolumn{2}{c}{\textsf{Mroibvol}} 
     & \multicolumn{2}{c}{\textsf{Mroibtrd}}
     & \multicolumn{2}{c}{\textsf{Mroibvol}} 
     & \multicolumn{2}{c}{\textsf{Mroibtrd}}\\ 
     \cmidrule(lr){2-3}\cmidrule(lr){4-5}\cmidrule(lr){6-7}\cmidrule(lr){8-9}
         & Coef & \textit{t}-stat & Coef & \textit{t}-stat & Coef & \textit{t}-stat & Coef & \textit{t}-stat \\
\midrule
        Intercept & $-$0.2072 & $-$27.82 & $-$0.1472 & $-$19.54 & $-$0.3968 & $-$33.48 & $-$0.2280 & $-$23.28 \\ 
        \textsf{Mroib}($w-1$) & 0.0983 & 57.30 & 0.2237 & 84.06 & 0.1729 & 80.38 & 0.2978 & 128.10 \\ 
        \textsf{Ret}($w-1$) & $-$0.3567 & $-$25.85 & $-$0.4100 & $-$28.71 & $-$0.5704 & $-$30.34 & $-$0.5363 & $-$31.47 \\ 
        \textsf{Ret}($m-1$)  & $-$0.0933 & $-$12.49 & $-$0.1023 & $-$13.87 & $-$0.1475 & $-$14.92 & $-$0.0823 & $-$9.47 \\ 
        \textsf{Ret}($m-7,m-2$) & $-$0.0225 & $-$6.72 & $-$0.0122 & $-$4.34 & $-$0.0296 & $-$6.86 & 0.0024 & 0.75 \\ 
        \textsf{Lmto} & 0.0000 & 2.14 & 0.0001 & 4.82 & 0.0002 & 6.04 & 0.0001 & 6.40 \\ 
        \textsf{Lvol} & 0.4199 & 6.59 & 0.5800 & 9.73 & 1.2780 & 13.83 & 0.8878 & 11.30 \\ 
        \textsf{Size} & 0.0151 & 18.36 & 0.0140 & 14.52 & 0.0282 & 26.06 & 0.0220 & 19.40 \\ 
        \textsf{Lbm} & $-$0.0085 & $-$10.13 & $-$0.0099 & $-$12.57 & $-$0.0173 & $-$18.01 & $-$0.0186 & $-$21.69 \\ 
        Adj.$R^2$ & 1.41\% & ~ & 6.14\% & ~ & 4.28\% & ~ & 10.68\% \\ 
        \bottomrule
    \end{tabular}}
        \begin{minipage}{15cm}
        \vspace{0.1cm}
        \footnotesize  *\textit{Numbers used in this discussion pertain to \textsf{Mroibvol}. Using \textsf{Mroibtrd} instead does not fundamentally change the interpretation.}
        \end{minipage}
\end{table}
\restoregeometry

\newgeometry{left=2cm,right=2cm,top=2cm,bottom=2cm}
\begin{table}[!h]
\caption{\textbf{Predicting Next-Week Returns Using Marketable Retail Order Imbalances}\\
\textbf{Description}: This table displays results analogous to Table III of BJZZ. Specifically, BJZZ examine the predictive power of order imbalances on future returns through the following \citetalias{FamaMcBeth1973} two-stage estimation: $\textsf{Ret}(i,w) = c_0(w) + c_1(w)\textsf{Mroib}(i,w-1) + c_2(w)'\textsf{CTRL}(i,w-1) +u_2(i,w)$. The (second-stage) standard errors' estimates are calculated using \citetalias{NW1987} with five lags. In Panel (a), we report our replication of their original findings. In Panels (b), (c), and (d), we revisit them using a more recent period and the alternative QMP method. To save space, we only report results based on bid-ask returns. \\ 
\textbf{Interpretation*}: 
BJZZ's original results (Panel (a)) indicate that past-week ROIs can predict future returns in the same direction. Our recent-period results based on the BJZZ method reveal a considerably weaker predictive power compared to BJZZ's original findings in the 2010-2015 period ((c) vs. (a)). 
Specifically, the coefficient estimate of $\textsf{Mroibvol}(w-1)$ and its $t$-stat are 34\% (0.0006 vs. 0.0009) and 48\% (7.95 vs. 15.14) lower, respectively, and the corresponding economic magnitude decreases from 11.16 bps to 6.02 bps per week. This finding holds when using QMP but to a lesser extent (Panel (d) vs. (b)). Comparing the BJZZ and QMP methods, during the 2010-2015 period, predictability holds with the same order of economic magnitude for both methods ((b) vs. (a)). In 2016-2021, however, QMP tends to reinforce the evidence for predictability ((d) vs. (c)). For example, the coefficient estimate on $\textsf{Mroibvol}(w-1)$ and its $t$-stat are 63\% and 28\% higher, respectively.
}
\label{tab:BJZZTAB3}
\centering
\scalebox{0.9}{
    \centering
    \begin{tabular}{lrrrrrrrr}
    \toprule
     & \multicolumn{4}{c}{\textit{(a) BJZZ 2010-2015}}  & \multicolumn{4}{c}{\textit{(b) QMP 2010-2015}} \\
     \cmidrule(lr){2-5} \cmidrule(lr){6-9}
     & \multicolumn{2}{c}{\textsf{Mroibvol}} 
     & \multicolumn{2}{c}{\textsf{Mroibtrd}}
     & \multicolumn{2}{c}{\textsf{Mroibvol}} 
     & \multicolumn{2}{c}{\textsf{Mroibtrd}}\\ 
                  \cmidrule(lr){2-3}\cmidrule(lr){4-5}\cmidrule(lr){6-7}\cmidrule(lr){8-9}
         & Coef & \textit{t}-stat & Coef & \textit{t}-stat & Coef & \textit{t}-stat & Coef & \textit{t}-stat \\ 
\midrule
        Intercept & 0.0033 & 2.24 & 0.0033 & 2.23 & 0.0034 & 2.33 & 0.0033 & 2.26 \\ 
        \textsf{Mroib}($w-1$)  & 0.0009 & 15.14 & 0.0008 & 11.93 & 0.0010 & 14.84 & 0.0009 & 12.13 \\ 
        \textsf{Ret}($w-1$)  & $-$0.0172 & $-$5.45 & $-$0.0174 & $-$5.50 & $-$0.0181 & $-$5.73 & $-$0.0183 & $-$5.79 \\ 
        \textsf{Ret}($m-1$) & $-$0.0001 & $-$0.03 & $-$0.0001 & $-$0.07 & $-$0.0001 & $-$0.04 & $-$0.0002 & $-$0.13 \\ 
        \textsf{Ret}($m-7,m-2$) & 0.0007 & 1.07 & 0.0007 & 1.05 & 0.0007 & 1.01 & 0.0007 & 0.96 \\ 
        \textsf{Lmto} & $-$0.0000 & $-$2.79 & $-$0.0000 & $-$2.76 & $-$0.0000 & $-$2.83 & $-$0.0000 & $-$2.76 \\ 
        \textsf{Lvol} & $-$0.0133 & $-$0.81 & $-$0.0130 & $-$0.79 & $-$0.0132 & $-$0.80 & $-$0.0124 & $-$0.75 \\ 
        \textsf{Size} & $-$0.0000 & $-$0.11 & $-$0.0000 & $-$0.17 & $-$0.0000 & $-$0.17 & $-$0.0000 & $-$0.18 \\ 
        \textsf{Lbm} & 0.0002 & 1.12 & 0.0002 & 1.10 & 0.0002 & 1.10 & 0.0002 & 1.05 \\ 
        Adj.$R^2$ & 3.75\% & ~ & 3.74\% & ~ & 3.76\% & ~ & 3.75\% & ~ \\ 
        IQR & 1.1950 & ~ & 1.2279 & ~ & 1.2014 & ~ & 1.1984 & ~ \\ 
        IQR w. ret. diff  & 0.1116\% & ~ & 0.0977\% & ~ & 0.1210\% & ~ & 0.1031\% & ~ \\ 
        \midrule
     & \multicolumn{4}{c}{\textit{(c) BJZZ 2016-2021}}  & \multicolumn{4}{c}{\textit{(d) QMP 2016-2021}} \\
     \cmidrule(lr){2-5} \cmidrule(lr){6-9}
     & \multicolumn{2}{c}{\textsf{Mroibvol}} 
     & \multicolumn{2}{c}{\textsf{Mroibtrd}}
     & \multicolumn{2}{c}{\textsf{Mroibvol}} 
     & \multicolumn{2}{c}{\textsf{Mroibtrd}}\\ 
                       \cmidrule(lr){2-3}\cmidrule(lr){4-5}\cmidrule(lr){6-7}\cmidrule(lr){8-9}
         & Coef & \textit{t}-stat & Coef. & \textit{t}-stat & Coef & \textit{t}-stat & Coef & \textit{t}-stat \\
\midrule
        Intercept & 0.0041 & 2.24 & 0.0042 & 2.25 & 0.0044 & 2.38 & 0.0042 & 2.26 \\ 
        \textsf{Mroib}($w-1$) & 0.0006 & 7.95 & 0.0007 & 4.86 & 0.0010 & 10.17 & 0.0007 & 4.81 \\ 
        \textsf{Ret}($w-1$) & $-$0.0138 & $-$3.73 & $-$0.0138 & $-$3.74 & $-$0.0141 & $-$3.82 & $-$0.0140 & $-$3.80 \\ 
        \textsf{Ret}($m-1$) & $-$0.0019 & $-$0.98 & $-$0.0019 & $-$0.97 & $-$0.0018 & $-$0.91 & $-$0.0019 & $-$0.97 \\ 
        \textsf{Ret}($m-7,m-2$) & $-$0.0001 & $-$0.19 & $-$0.0001 & $-$0.20 & $-$0.0001 & $-$0.18 & $-$0.0001 & $-$0.23 \\ 
        \textsf{Lmto} & $-$0.0000 & $-$0.81 & $-$0.0000 & $-$0.82 & $-$0.0000 & $-$0.84 & $-$0.0000 & $-$0.83 \\ 
        \textsf{Lvol} & 0.0125 & 0.73 & 0.0123 & 0.72 & 0.0110 & 0.65 & 0.0121 & 0.71 \\ 
        \textsf{Size} & $-$0.0002 & $-$1.08 & $-$0.0002 & $-$1.12 & $-$0.0002 & $-$1.14 & $-$0.0002 & $-$1.10 \\ 
        \textsf{Lbm} & 0.0001 & 0.62 & 0.0001 & 0.63 & 0.0001 & 0.73 & 0.0001 & 0.73 \\ 
        Adj.$R^2$  & 4.35\% & ~ & 4.37\% & ~ & 4.38\% & ~ & 4.39\% & ~ \\ 
        IQR & 0.9863 & ~ & 0.8114 & ~ & 1.0746 & ~ & 0.9831 & ~ \\ 
        IQR w. ret. diff  & 0.0602\% & ~ & 0.0548\% & ~ & 0.1069\% & ~ & 0.0699\% \\ 
        \bottomrule
    \end{tabular}}
        \begin{minipage}{15cm}
        \vspace{0.1cm}
        \footnotesize  *\textit{Numbers used in this discussion pertain to \textsf{Mroibvol}. Using \textsf{Mroibtrd} instead does not fundamentally change the interpretation.}
        \end{minipage}
\end{table}
\restoregeometry

\newgeometry{left=2cm,right=2cm,top=2cm,bottom=2cm}
\begin{table}[!h]
\caption{\textbf{Marketable Retail Return Predictability Within Subgroups} \\
\textbf{Description}: This table displays results analogous to Table IV of BJZZ. Specifically, BJZZ analyze the predictive power of order imbalances on future returns conditional on three firms’ characteristics: market capitalization, share price, and turnover. They estimate variants of specifications~\eqref{eq:Reg_PREDICT}, where all coefficients are allowed to be different within each subgroup. Standard errors’ estimates are calculated using \citetalias{NW1987} with five lags. In Panel (a), we report our replication of their original findings. In Panels (b), (c), and (d), we revisit them using a more recent period and the alternative QMP method. We only report results based on \textsf{Mroibvol} to save space.\\
\textbf{Interpretation}: 
BJZZ’s original results (Panel (a)) indicate that the predictability exists for all market-cap, share price and turnover groups; and that the predictability is stronger for small-cap, low-price, and low-turnover stocks. Comparing periods with the BJZZ method ((c) vs. (a)) reveals that the original conclusions tend to weaken or disappear in 2016-2021. For example, the economic magnitude associated with the predictability of small-cap stocks decreases from 20.5 bps to 9.8 bps per year. Additionally, the statistically significant (1\% level) predictive power for big-cap and high-price stocks that existed in 2010-2015 completely disappears in 2016-2021. When we compare periods using the QMP method instead ((d) vs. (b)), the predictive power still weakens or disappears in the recent period, but to a lesser extent. Contrasting results between methods show that both yield similar results in 2010-2015 ((b) vs. (a)), but important differences arise in the 2016-2021 period, with the QMP method suggesting stronger predictability for most subgroups ((d) vs. (c)). For instance, the economic magnitude associated with the predictability of returns on small-cap stocks is approximately twice as large with QMP (19.7 bps vs. 9.8 bps).}
\label{tab:BJZZTAB4}
\centering
\scalebox{0.9}{
    \centering
    \begin{tabular}{lrrrrrrrr}
    \toprule
    & \multicolumn{4}{c}{\textit{(a) BJZZ 2010-2015}}  & \multicolumn{4}{c}{\textit{(b) QMP 2010-2015}} \\
     \cmidrule(lr){2-5} \cmidrule(lr){6-9}
         & Coef & \textit{t}-stat & IQR &  W.R. Diff. & Coef & \textit{t}-stat & IQR & W.R. Diff. \\ 
     \midrule
        \multicolumn{9}{l}{\textit{Market-Cap Subgroups}} \\
        Small & 0.0013 & 13.87 & 1.6010 & 0.205\% & 0.0014 & 14.47 & 1.6209 & 0.223\% \\ 
        Medium & 0.0005 & 6.70 & 1.2386 & 0.068\% & 0.0005 & 5.73 & 1.2353 & 0.065\% \\ 
        Big & 0.0003 & 3.79 & 0.8746 & 0.028\% & 0.0004 & 4.43 & 0.8804 & 0.037\% \\ 
        \multicolumn{9}{l}{\textit{Share-Price Subgroups}} \\
         Low & 0.0015 & 13.40 & 1.4088 & 0.205\% & 0.0016 & 13.39 & 1.4106 & 0.219\% \\ 
        Medium & 0.0006 & 7.76 & 1.2672 & 0.074\% & 0.0006 & 7.89 & 1.2606 & 0.080\% \\ 
        High & 0.0002 & 3.37 & 0.9495 & 0.023\% & 0.0003 & 4.38 & 0.9703 & 0.032\% \\ 
        \multicolumn{9}{l}{\textit{Turnover Subgroups}} \\
        Low & 0.0010 & 14.99 & 1.7156 & 0.176\% & 0.0011 & 15.46 & 1.7491 & 0.193\% \\ 
        Medium & 0.0008 & 8.32 & 1.1589 & 0.090\% & 0.0009 & 9.18 & 1.1550 & 0.101\% \\ 
        High & 0.0009 & 5.19 & 0.8681 & 0.074\% & 0.0009 & 4.84 & 0.8661 & 0.075\% \\
    \midrule
     & \multicolumn{4}{c}{\textit{(c) BJZZ 2016-2021}}  & \multicolumn{4}{c}{\textit{(d) QMP 2016-2021}} \\
     \cmidrule(lr){2-5} \cmidrule(lr){6-9}
         & Coef & \textit{t}-stat & IQR &  W.R. Diff. & Coef & \textit{t}-stat & IQR & W.R. Diff. \\ 
     \midrule
        \multicolumn{9}{l}{\textit{Market-Cap Subgroups}} \\
        Small & 0.0007 & 6.44 & 1.3347 & 0.098\% & 0.0014 & 10.35 & 1.4216 & 0.197\% \\ 
        Medium & 0.0004 & 4.03 & 1.0890 & 0.045\% & 0.0005 & 4.30 & 1.1385 & 0.058\% \\ 
        Big & 0.0002 & 1.54 & 0.6799 & 0.014\% & 0.0001 & 0.55 & 0.7859 & 0.006\% \\ 
        \multicolumn{9}{l}{\textit{Share-Price Subgroups}} \\
        Low & 0.0012 & 7.79 & 1.1461 & 0.133\% & 0.0018 & 10.63 & 1.2062 & 0.216\% \\ 
        Medium & 0.0002 & 2.31 & 1.1309 & 0.025\% & 0.0004 & 3.60 & 1.1889 & 0.047\% \\ 
        High & 0.0000 & 0.06 & 0.7410 & 0.000\% & 0.0003 & 3.13 & 0.8662 & 0.029\% \\ 
        \multicolumn{9}{l}{\textit{Turnover Subgroups}} \\
        Low & 0.0006 & 7.48 & 1.4643 & 0.082\% & 0.0011 & 12.41 & 1.5670 & 0.176\% \\ 
        Medium & 0.0007 & 4.72 & 0.9304 & 0.064\% & 0.0008 & 5.28 & 1.0188 & 0.084\% \\ 
        High & 0.0007 & 2.69 & 0.7346 & 0.049\% & 0.0010 & 3.85 & 0.7951 & 0.078\% \\
        \bottomrule
    \end{tabular}}
\end{table}%
\restoregeometry

\newgeometry{left=2cm,right=2cm,top=2cm,bottom=2cm}
\begin{table}[!h]
\caption{\textbf{Predicting Returns $k$ Weeks Ahead} \\
\textbf{Description}: This table displays results analogous to Table V of BJZZ. Specifically, BJZZ analyze the predictive power of marketable retail order imbalances at horizons longer than one week through the following \citetalias{FamaMcBeth1973} two-stage estimation: $\textsf{Ret}(i,w+k) = c_0(w) + c_1(w) \textsf{Mroib}(i,w) + c_2(w)'\textsf{CTRL}(i,w) +u_3(i,w+k)$. The (second-stage) standard errors' estimates are calculated using \citetalias{NW1987} with five lags, and $\textsf{Ret}(i,w+k)$ represents the \emph{one-week period} return $k$ week ahead, rather than a cumulative return over $k$ week. In Panel (a), we report our replication of their original findings. In Panels (b), (c), and (d), we revisit them using a more recent period and the alternative QMP method. To save space, we only report results based on bid-ask returns. \\ 
\textbf{Interpretation*}: BJZZ's original results (Panel (a)) indicate that retail order imbalances can predict future returns up to six to eight weeks ahead and that the predictive power generally decreases monotonically with the horizon. Comparing periods with the BJZZ method ((c) vs. (a)) shows that the horizon of predictability shortens in the recent period. For instance, in 2010-2015, ROI can predict returns up to eight weeks ahead (\eg, eight-week \textsf{Mroibvol} coefficient of 0.0002 with a $t$-stat of 3.96), whereas in 2016-2021, the predictive significance starts to weaken at four weeks and beyond. Indeed, the 4 and 6 weeks \textsf{Mroibvol} coefficients of 0.0002 and 0.0002 are significant at the 5\% level only, and the eight weeks coefficient of 0.0001 is significant at the 10\% level only. When comparing periods using the QMP method ((d) vs. (b)), this interpretation holds true but to a lesser extent, as \textsf{Mroibvol}'s predictive power loses statistical significance rather after six weeks than four weeks. Comparing methods ((b) vs. (a) and (d) vs. (c)), both lead to similar conclusions and economic magnitudes in 2010-2015, but notable differences arise in the 2016-2021 period, with stronger and more significant predictive coefficients for all horizons when using the QMP method. For example, the two weeks \textsf{Mroibvol} coefficient and $t$-stat are respectively 0.0004 and 4.19 for the QMP method compared to 0.0003 and 3.46 for the BJZZ method; and the four-week \textsf{Mroibvol} coefficient and $t$-stat are respectively 0.0004 and 3.91 for the QMP method compared to 0.0002 and 2.02 for the BJZZ method.}
\label{tab:BJZZTAB5}
\centering
\scalebox{0.9}{
    \centering
    \begin{tabular}{lcccccccc}
    \toprule
     & \multicolumn{4}{c}{\textit{(a) BJZZ 2010-2015}}  & \multicolumn{4}{c}{\textit{(b) QMP 2010-2015}} \\
     \cmidrule(lr){2-5} \cmidrule(lr){6-9}
     & \multicolumn{2}{c}{\textsf{Mroibvol}} & \multicolumn{2}{c}{\textsf{Mroibtrd}}
     & \multicolumn{2}{c}{\textsf{Mroibvol}} & \multicolumn{2}{c}{\textsf{Mroibtrd}}\\ 
     \cmidrule(lr){2-3}\cmidrule(lr){4-5}\cmidrule(lr){6-7}\cmidrule(lr){8-9}
         & Coef & \textit{t}-stat & Coef. & \textit{t}-stat & Coef & \textit{t}-stat & Coef & \textit{t}-stat \\ 
\midrule
        1 week & 0.0009 & 15.14 & 0.0008 & 11.93 & 0.0010 & 14.84 & 0.0009 & 12.13 \\ 
        2 weeks & 0.0006 & 9.48 & 0.0005 & 7.77 & 0.0006 & 9.00 & 0.0005 & 7.09 \\ 
        4 weeks & 0.0003 & 5.64 & 0.0003 & 5.40 & 0.0003 & 5.52 & 0.0003 & 5.59 \\ 
        6 weeks & 0.0003 & 4.53 & 0.0002 & 3.31 & 0.0003 & 4.64 & 0.0002 & 3.67 \\ 
        8 weeks & 0.0002 & 3.96 & 0.0002 & 2.43 & 0.0002 & 2.51 & 0.0001 & 1.96 \\ 
        10 weeks & 0.0000 & 0.78 & -0.0001 & -0.87 & 0.0000 & 0.21 & -0.0001 & -0.87 \\ 
        12 weeks & 0.0001 & 2.48 & 0.0002 & 2.68 & 0.0002 & 2.92 & 0.0002 & 3.30 \\ 
        \midrule
     & \multicolumn{4}{c}{\textit{(c) BJZZ 2016-2021}}  & \multicolumn{4}{c}{\textit{(d) QMP 2016-2021}} \\
     \cmidrule(lr){2-5} \cmidrule(lr){6-9}
     & \multicolumn{2}{c}{\textsf{Mroibvol}} & \multicolumn{2}{c}{\textsf{Mroibtrd}}
     & \multicolumn{2}{c}{\textsf{Mroibvol}} & \multicolumn{2}{c}{\textsf{Mroibtrd}}\\ 
     \cmidrule(lr){2-3}\cmidrule(lr){4-5}\cmidrule(lr){6-7}\cmidrule(lr){8-9}
         & Coef & \textit{t}-stat & Coef & \textit{t}-stat & Coef & \textit{t}-stat & Coef & \textit{t}-stat \\ 
\midrule
        1 week & 0.0006 & 7.95 & 0.0007 & 4.86 & 0.0010 & 10.17 & 0.0007 & 4.81 \\ 
        2 weeks & 0.0003 & 3.46 & 0.0003 & 2.27 & 0.0004 & 4.19 & 0.0002 & 1.54 \\ 
        4 weeks & 0.0002 & 2.02 & 0.0003 & 1.79 & 0.0004 & 3.91 & 0.0003 & 2.03 \\ 
        6 weeks & 0.0002 & 2.10 & 0.0002 & 1.61 & 0.0002 & 2.20 & 0.0001 & 0.99 \\ 
        8 weeks & 0.0001 & 1.79 & 0.0003 & 1.97 & 0.0002 & 1.81 & 0.0001 & 1.01 \\ 
        10 weeks & 0.0002 & 2.74 & 0.0003 & 2.49 & 0.0003 & 3.61 & 0.0003 & 2.47 \\ 
        12 weeks & 0.0002 & 2.55 & 0.0003 & 2.20 & 0.0004 & 3.73 & 0.0003 & 2.21 \\ 
    \bottomrule
    \end{tabular}}
        \begin{minipage}{15cm}
        \vspace{0.1cm}
        \footnotesize  *\textit{Numbers used in this discussion pertain to \textsf{Mroibvol}. Using \textsf{Mroibtrd} instead does not fundamentally change the interpretation, with perhaps the exception of Panel (d) where the \textsf{Mroibtrd} coefficients appear to have a weaker significance level at all horizon longer than one week.}
        \end{minipage}
\end{table}
\restoregeometry

\newgeometry{left=1.3cm,right=1.3cm,top=2cm,bottom=2cm}
\begin{landscape}
\begin{table}[!h]
\caption{\textbf{Long-Short Strategy Returns Based on Marketable Retail Order Imbalances}\\
\textbf{Description}: This table displays results analogous to Table VI of BJZZ. BJZZ analyze whether marketable retail order imbalances can be used as a signal to form a profitable trading strategy. They form quintile portfolios based on the previous week’s average order imbalance and construct long-short portfolios where the stocks in the highest (lowest) order imbalance quintile are bought (shorted). The performance of the portfolios is assessed in terms of raw returns and alpha against the \citet{FF1993} three-factor model, and for horizons up to 12 weeks. The returns are value-weighted using the previous month-end market cap. The standard errors are adjusted using \citet{HansenHodrick} with a dynamic number of lags as a function of the horizon. In Panel (a), we report our replication of their original findings. In Panels (b), (c), and (d), we revisit them using a more recent period and the alternative QMP method. We only report results based on \textsf{Mroibvol} and alphas to save space. \\
\textbf{Interpretation}: BJZZ’s original results (Panel (a)) indicate that a long-short strategy based on marketable retail order imbalances generates statistically positive alphas at horizons from one to 12 weeks, and that results are more pronounced with a universe of small-cap stocks only. In the 2016-2021 period, this strategy ceases to be profitable when we consider all stocks. Indeed, alphas of the recent-period panels (c) and (d) are no longer statistically significant, regardless of the horizon or method. When considering small-cap stocks only, alphas remain statistically positive but experience a notable decline. Specifically, based on BJZZ samples ((c) vs. (a)), at all horizons, small-sample alphas and their $t$-stat are much lower compared to the 2010-2015 period, with differences surpassing 65\% for one- and two-week horizons (0.143\% vs. 0.437\% and 0.177\% vs. 0.613\%, respectively). Moreover, at horizons of 8 weeks and beyond, the recent-period small cap alphas are no longer significant at the 10\% level. Based on QMP samples ((d) vs. (b)), the small-cap alphas also experience an important but more moderate decline. Contrasting methods, in general, both lead to similar conclusions and economic magnitudes in both periods ((b) vs. (a) and (d) vs. (c))---with some notable differences in the recent period, such as significantly stronger one-week small cap alpha for the QMP method (0.295\% vs. 0.143\%).}
\label{tab:BJZZTAB6}
\centering
\scalebox{0.8}{
    \centering
    \begin{tabular}{lrrrrrrrrrrrrrrrr}
   \toprule
     & \multicolumn{8}{c}{\textit{(a) BJZZ 2010-2015}}  & \multicolumn{8}{c}{\textit{(b) QMP 2010-2015}} \\
     \cmidrule(lr){2-9} \cmidrule(lr){10-17}
     & \multicolumn{2}{c}{All Stocks} & \multicolumn{2}{c}{Small}
     & \multicolumn{2}{c}{Medium} & \multicolumn{2}{c}{Big} & \multicolumn{2}{c}{All Stocks} & \multicolumn{2}{c}{Small}
     & \multicolumn{2}{c}{Medium} & \multicolumn{2}{c}{Big} \\ 
     \cmidrule(lr){2-3} \cmidrule(lr){4-5} \cmidrule(lr){6-7} \cmidrule(lr){8-9} \cmidrule(lr){10-11} \cmidrule(lr){12-13} \cmidrule(lr){14-15} \cmidrule(lr){16-17}
         & Alpha & \textit{t}-stat & Alpha & \textit{t}-stat & Alpha & \textit{t}-stat & Alpha & \textit{t}-stat & Alpha & \textit{t}-stat & Alpha & \textit{t}-stat & Alpha & \textit{t}-stat & Alpha & \textit{t}-stat \\ 
\midrule
        1 week & 0.083\% & 2.77 & 0.437\% & 10.39 & 0.175\% & 5.59 & 0.051\% & 1.52 & 0.072\% & 2.18 & 0.411\% & 9.99 & 0.180\% & 5.25 & 0.033\% & 1.02 \\ 
        2 weeks & 0.090\% & 1.81 & 0.613\% & 8.68 & 0.270\% & 5.01 & 0.052\% & 1.04 & 0.087\% & 1.72 & 0.607\% & 8.11 & 0.257\% & 4.63 & 0.047\% & 0.89 \\ 
        4 weeks & 0.167\% & 2.04 & 0.852\% & 7.15 & 0.377\% & 4.46 & 0.125\% & 1.54 & 0.182\% & 1.97 & 0.769\% & 7.30 & 0.366\% & 4.23 & 0.124\% & 1.35 \\ 
        6 weeks & 0.285\% & 2.56 & 0.909\% & 6.54 & 0.471\% & 3.88 & 0.193\% & 1.76 & 0.272\% & 2.26 & 0.871\% & 5.85 & 0.453\% & 3.83 & 0.214\% & 1.80 \\ 
        8 weeks & 0.412\% & 2.57 & 0.992\% & 4.96 & 0.523\% & 3.07 & 0.297\% & 2.03 & 0.341\% & 2.09 & 0.983\% & 5.28 & 0.514\% & 3.11 & 0.253\% & 1.61 \\ 
        10 weeks & 0.373\% & 1.73 & 0.905\% & 3.68 & 0.406\% & 2.58 & 0.263\% & 1.40 & 0.226\% & 1.02 & 0.893\% & 3.96 & 0.373\% & 2.23 & 0.101\% & 0.53 \\ 
        12 weeks & 0.564\% & 2.07 & 0.988\% & 4.02 & 0.364\% & 2.05 & 0.416\% & 1.64 & 0.384\% & 1.32 & 0.878\% & 3.60 & 0.331\% & 1.83 & 0.203\% & 0.81 \\ 
        \midrule
     & \multicolumn{8}{c}{\textit{(c) BJZZ 2016-2021}}  & \multicolumn{8}{c}{\textit{(d) QMP 2016-2021}} \\
     \cmidrule(lr){2-9} \cmidrule(lr){10-17}
     & \multicolumn{2}{c}{All Stocks} & \multicolumn{2}{c}{Small}
     & \multicolumn{2}{c}{Medium} & \multicolumn{2}{c}{Big} & \multicolumn{2}{c}{All Stocks} & \multicolumn{2}{c}{Small}
     & \multicolumn{2}{c}{Medium} & \multicolumn{2}{c}{Big} \\ 
     \cmidrule(lr){2-3} \cmidrule(lr){4-5} \cmidrule(lr){6-7} \cmidrule(lr){8-9} \cmidrule(lr){10-11} \cmidrule(lr){12-13} \cmidrule(lr){14-15} \cmidrule(lr){16-17}
         & Alpha & \textit{t}-stat & Alpha & \textit{t}-stat & Alpha & \textit{t}-stat & Alpha & \textit{t}-stat & Alpha & \textit{t}-stat & Alpha & \textit{t}-stat & Alpha & \textit{t}-stat & Alpha & \textit{t}-stat \\ 
 \midrule
        1 week & $-$0.021\% & $-$0.48 & 0.143\% & 3.16 & 0.067\% & 1.61 & $-$0.042\% & $-$0.91 & $-$0.009\% & $-$0.17 & 0.295\% & 5.11 & 0.098\% & 1.97 & $-$0.043\% & $-$1.01 \\ 
        2 weeks & 0.004\% & 0.07 & 0.177\% & 2.54 & 0.025\% & 0.41 & $-$0.075\% & $-$1.10 & $-$0.060\% & $-$0.92 & 0.347\% & 3.44 & 0.116\% & 1.56 & $-$0.100\% & $-$1.32 \\ 
        4 weeks & $-$0.018\% & $-$0.20 & 0.381\% & 2.75 & $-$0.013\% & $-$0.14 & $-$0.114\% & $-$1.01 & $-$0.145\% & $-$1.18 & 0.507\% & 3.06 & 0.074\% & 0.70 & $-$0.161\% & $-$1.15 \\ 
        6 weeks & $-$0.126\% & $-$1.02 & 0.424\% & 2.56 & $-$0.128\% & $-$1.24 & $-$0.211\% & $-$1.62 & $-$0.254\% & $-$1.52 & 0.644\% & 3.51 & $-$0.020\% & $-$0.13 & $-$0.336\% & $-$2.08 \\ 
        8 weeks & $-$0.241\% & $-$1.36 & 0.171\% & 0.73 & $-$0.199\% & $-$1.52 & $-$0.260\% & $-$1.34 & $-$0.353\% & $-$1.57 & 0.526\% & 2.62 & $-$0.222\% & $-$1.44 & $-$0.420\% & $-$1.97 \\ 
        10 weeks & $-$0.243\% & $-$1.00 & 0.097\% & 0.38 & $-$0.221\% & $-$1.41 & $-$0.187\% & $-$0.70 & $-$0.361\% & $-$1.28 & 0.293\% & 1.63 & $-$0.388\% & $-$2.08 & $-$0.415\% & $-$1.60 \\ 
        12 weeks & $-$0.255\% & $-$0.86 & $-$0.014\% & $-$0.04 & $-$0.284\% & $-$1.82 & $-$0.214\% & $-$0.77 & $-$0.442\% & $-$1.23 & 0.083\% & 0.49 & $-$0.524\% & $-$2.36 & $-$0.458\% & $-$1.35 \\
        \bottomrule
    \end{tabular}}
\end{table}
\end{landscape}
\restoregeometry

\newgeometry{left=2cm,right=2cm,top=2cm,bottom=2cm}
\begin{table}[!h]
\caption{\textbf{Predictability Decomposition} \\
\textbf{Description}: This table displays results analogous to Table VII of BJZZ. BJZZ explore three alternative hypotheses that could elucidate why marketable retail order imbalance can predict future returns. The first hypothesis hinges on the persistence of order flows; the second hypothesis relies on the contrarian trading behavior exhibited by retail investors; and the third hypothesis posits that retail investors may accurately predict the direction of future returns because they possess valuable information about the firm. To test these hypotheses, BJZZ adopt a two-stage decomposition. First, they decompose their retail marketable order imbalance variable in three parts as $\textsf{Mroib}(i,w) = \widehat{\textsf{Mroib}^{persistence}_{i,w}} + \widehat{\textsf{Mroib}^{contrarian}_{i,w}} + \widehat{\textsf{Mroib}^{other}_{i,w}}$. Then, they estimate~\eqref{eq:Reg_PREDICT} where $\textsf{Mroib}(i,w)$ is replaced by its three components. Standard errors' estimates are calculated using \citetalias{NW1987} with five lags. In Panel (a), we report our replication of their original findings. In Panels (b), (c), and (d), we revisit them using a more recent period and the alternative QMP method. To save space, we only report results on the three components from the second-stage decomposition, and based on bid-ask returns. \\
\textbf{Interpretation*}: BJZZ's original results indicate that most of the predictability primarily comes from the persistence (PERS) and residual (OTHER) components of retail order imbalance---the latter aligning with the third hypothesis described above. They also show that the contrarian trading component (CONT) lacks statistical significance. In the recent period, the persistence and residual components, though still significant, seriously weaken. For example, based on the BJZZ samples ((c) vs. (a)), the significance of PERS drops from 1 to 10\%, and its economic magnitude decreases from 0.0692\% to 0.0241\%. If we rather consider the QMP samples ((d) vs. (b)), we observe a similar trend, albeit less pronounced. Finally, regardless of the method used, the CONT component remain insignificant in 2016-2021. When comparing methods ((b) vs. (a) and (d) vs. (c)), results mostly align in 2010-2015. However, in 2016-2021, the persistence and residual components show greater statistical significance and larger economic magnitudes when we rely on QMP (\eg, PERS and OTHER based on \textsf{Mroibvol} in Panel (d) have economic magnitudes of 0.0534\% and 0.0941\%, respectively, compared to 0.0241\% and 0.0580\% in Panel (c)).}
\label{tab:BJZZTAB7}
\centering
\scalebox{0.8}{
    \centering
    \begin{tabular}{lrrrrrrrr}
    \toprule
     & \multicolumn{4}{c}{\textit{(a) BJZZ 2010-2015}}  & \multicolumn{4}{c}{\textit{(b) QMP 2010-2015}} \\
     \cmidrule(lr){2-5} \cmidrule(lr){6-9}
     & \multicolumn{2}{c}{\textsf{Mroibvol} } & \multicolumn{2}{c}{\textsf{Mroibtrd}}
     & \multicolumn{2}{c}{\textsf{Mroibvol} } & \multicolumn{2}{c}{\textsf{Mroibtrd}}\\ 
                  \cmidrule(lr){2-3}\cmidrule(lr){4-5}\cmidrule(lr){6-7}\cmidrule(lr){8-9}
         & Coef & \textit{t}-stat & Coef & \textit{t}-stat & Coef & \textit{t}-stat & Coef & \textit{t}-stat \\ 
\midrule
        PERS & 0.0030 & 8.14 & 0.0019 & 7.35 & 0.0026 & 8.32 & 0.0018 & 6.89 \\ 
        CONT & $-$0.0114 & $-$0.42 & $-$0.0227 & $-$0.82 & 0.0057 & 0.69 & $-$0.0105 & $-$0.27 \\ 
        OTHER & 0.0008 & 13.02 & 0.0006 & 9.73 & 0.0009 & 13.28 & 0.0007 & 10.47 \\ 
        & \multicolumn{2}{c}{\textsf{Mroibvol} } & \multicolumn{2}{c}{\textsf{Mroibtrd}}
     & \multicolumn{2}{c}{\textsf{Mroibvol} } & \multicolumn{2}{c}{\textsf{Mroibtrd}}\\ 
                  \cmidrule(lr){2-3}\cmidrule(lr){4-5}\cmidrule(lr){6-7}\cmidrule(lr){8-9}
        ~ & IQR & R. Diff & IQR & R. Diff & IQR & R. Diff & IQR & R. Diff \\ 
\midrule
        PERS & 0.2319 & 0.0692\% & 0.3305 & 0.0636\% & 0.2836 & 0.0745\% & 0.3466 & 0.0635\% \\ 
        CONT & 0.0358 & $-$0.0408\% & 0.0328 & $-$0.0743\% & 0.0401 & 0.0229\% & 0.0358 & $-$0.0375\% \\ 
        OTHER & 1.1260 & 0.0902\% & 1.1333 & 0.0734\% & 1.1190 & 0.0953\% & 1.0977 & 0.0788\% \\
    \midrule
     & \multicolumn{4}{c}{\textit{(c) BJZZ 2016-2021}}  & \multicolumn{4}{c}{\textit{(d) QMP 2016-2021}} \\
     \cmidrule(lr){2-5} \cmidrule(lr){6-9}
     & \multicolumn{2}{c}{\textsf{Mroibvol} } & \multicolumn{2}{c}{\textsf{Mroibtrd}}
     & \multicolumn{2}{c}{\textsf{Mroibvol} } & \multicolumn{2}{c}{\textsf{Mroibtrd}}\\ 
                  \cmidrule(lr){2-3}\cmidrule(lr){4-5}\cmidrule(lr){6-7}\cmidrule(lr){8-9}
         & Coef & \textit{t}-stat & Coef & \textit{t}-stat & Coef & \textit{t}-stat & Coef & \textit{t}-stat \\ 
\midrule
        PERS & 0.0026 & 1.80 & 0.0022 & 3.23 & 0.0029 & 4.64 & 0.0043 & 2.29 \\ 
        CONT & $-$0.0885 & $-$1.01 & $-$0.0074 & $-$0.13 & 0.0324 & 0.78 & 0.0013 & 2.45 \\ 
        OTHER & 0.0006 & 7.68 & 0.0006 & 4.89 & 0.0009 & 9.94 & $-$0.0271 & $-$0.72 \\ 
        & \multicolumn{2}{c}{\textsf{Mroibvol} } & \multicolumn{2}{c}{\textsf{Mroibtrd}}
     & \multicolumn{2}{c}{\textsf{Mroibvol} } & \multicolumn{2}{c}{\textsf{Mroibtrd}}\\ 
                  \cmidrule(lr){2-3}\cmidrule(lr){4-5}\cmidrule(lr){6-7}\cmidrule(lr){8-9}
        ~ & IQR & R. Diff & IQR & R. Diff & IQR & R. Diff & IQR & R. Diff \\ 
\midrule
        PERS & 0.0913 & 0.0241\% & 0.1810 & 0.0402\% & 0.1845 & 0.0534\% & 0.2970 & 0.0371\% \\ 
        CONT & 0.0164 & $-$0.1453\% & 0.0196 & $-$0.0146\% & 0.0266 & 0.0862\% & 0.0257 & $-$0.0697\% \\ 
        OTHER & 0.9640 & 0.0580\% & 0.7614 & 0.0480\% & 1.0331 & 0.0941\% & 0.8861 & 0.0579\% \\
        \bottomrule
    \end{tabular}}
        \begin{minipage}{15cm}
        \vspace{0.1cm}
        \footnotesize  *\textit{Numbers used in this discussion pertain to \textsf{Mroibvol}. Using \textsf{Mroibtrd} instead does not fundamentally change the interpretation.}
        \end{minipage}
\end{table}
\restoregeometry

\newgeometry{left=2cm,right=2cm,top=2cm,bottom=2cm}
\begin{table}[!h]
\caption{\textbf{Marketable Retail Order Imbalance and Contemporaneous Returns} \\
\textbf{Description}: 
This table displays results analogous to Table VIII in BJZZ. BJZZ explore the liquidity provision hypothesis, relying on the work of \citet{KanielSaarTitman2008} (KST). Specifically, they replicate Table III of KST. In this table, KST examine the past, contemporaneous, and future returns of intense buy and sell portfolios. Portfolios are constructed based on the previous week's net individual trading (NIT)---the KST equivalent measure of retail trading flows. KST's findings are threefold: (i) they observe typical contrarian trading behavior by retail investors; (ii) they show that retail trading can predict returns in the correct direction; and (iii) they obtain results in favor of the liquidity provision hypothesis. Table VIII of BJZZ corresponds to their replication results of table III of KST, where they construct portfolios using their own retail order imbalance measures, \textsf{Mroibvol} and \textsf{Mroibtrd}. In Panel (a), we report our replication of their original findings. In Panels (b), (c), and (d), we revisit them using a more recent period and the alternative QMP method. We only report results based on cumulative market-adjusted returns and intense buy and intense sell portfolios to save space. $^{**}$ and $^{*}$ indicate 1\% and 5\% level significance, respectively. We adjust \textit{t}-statistics using \citetalias{NW1987} with four lags. \\
\textbf{Interpretation}: 
BJZZ's original findings validate the first two observations of KST: the contrarian trading patterns of retail investors and the predictive power of order imbalance on future returns. However, BJZZ's results diverge from KST's third assertion concerning the liquidity provision hypothesis, with no supporting evidence found in BJZZ's findings, contrary to KST's. In the recent period, we observe changes in contrarian behaviors, primarily on the buying side. Retail investors still tend to buy stocks with negative returns but do not consistently sell stocks with positive returns. 
Regarding predictive power, the 2016-2021 results largely echo those of 2010-2015, with a noteworthy decline in significance specific to the buying side, dropping from 1 to 5\% or 10\% (see $k>0$ Intense Buying estimates in (c) vs. (a)). Finally, aligning with 2010-2015 results, we do not find evidence supporting the liquidity provision hypothesis in the recent period. Comparing methods ((b) vs. (a) and (d) vs. (c)), results align in 2010-2015, and slight variations emerge in the 2016-2021 period. For instance, the QMP method tends to provide more supporting evidence for contrarian behavior on the selling side compared to the BJZZ method.}
\label{tab:BJZZTAB8}
\centering
\scalebox{0.90}{
    \centering
    \begin{tabular}{lrrrrrrrr}
    \toprule
     & \multicolumn{4}{c}{\textit{(a) BJZZ 2010-2015}}  & \multicolumn{4}{c}{\textit{(b) QMP 2010-2015}} \\
     \cmidrule(lr){2-5} \cmidrule(lr){6-9}
     & \multicolumn{2}{c}{Intense Selling} & \multicolumn{2}{c}{Intense Buying}  & \multicolumn{2}{c}{Intense Selling} & \multicolumn{2}{c}{Intense Buying} \\ 
     \cmidrule(lr){2-3} \cmidrule(lr){4-5} \cmidrule(lr){6-7} \cmidrule(lr){8-9} 
         & Mean & \textit{t}-stat & Mean & \textit{t}-stat & Mean & \textit{t}-stat & Mean & \textit{t}-stat \\ 
  \midrule
        $k = -20$ & 0.0074$^{**}$ & 7.33 & $-$0.0166$^{**}$ & $-$19.30 & 0.0075$^{**}$ & 7.16 & $-$0.0154$^{**}$ & $-$18.32 \\ 
        $k = -15$ & 0.0071$^{**}$ & 9.34 & $-$0.0137$^{**}$ & $-$21.05 & 0.0074$^{**}$ & 9.49 & $-$0.0130$^{**}$ & $-$20.85 \\ 
        $k = -10$ & 0.0059$^{**}$ & 10.89 & $-$0.0103$^{**}$ & $-$20.67 & 0.0061$^{**}$ & 10.90 & $-$0.0101$^{**}$ & $-$21.55 \\ 
        $k = -5$ & 0.0039$^{**}$ & 12.56 & $-$0.0064$^{**}$ & $-$20.74 & 0.0040$^{**}$ & 11.46 & $-$0.0061$^{**}$ & $-$21.04 \\ 
        $k = 0$ & $-$0.0026$^{**}$ & $-$6.49 & 0.0021$^{**}$ & 5.44 & $-$0.0040$^{**}$ & $-$8.84 & 0.0048$^{**}$ & 10.84 \\ 
        $k = 5$ & $-$0.0017$^{**}$ & $-$6.58 & 0.0026$^{**}$ & 10.34 & $-$0.0019$^{**}$ & $-$7.31 & 0.0027$^{**}$ & 10.43 \\ 
        $k = 10$ & $-$0.0028$^{**}$ & $-$6.04 & 0.0039$^{**}$ & 9.09 & $-$0.0033$^{**}$ & $-$7.12 & 0.0041$^{**}$ & 9.12 \\ 
        $k = 15$ & $-$0.0039$^{**}$ & $-$6.10 & 0.0049$^{**}$ & 8.36 & $-$0.0047$^{**}$ & $-$7.45 & 0.0048$^{**}$ & 8.13 \\ 
        $k = 20$ & $-$0.0047$^{**}$ & $-$5.38 & 0.0052$^{**}$ & 6.27 & $-$0.0052$^{**}$ & $-$6.00 & 0.0051$^{**}$ & 6.36 \\ 
        \midrule
     & \multicolumn{4}{c}{\textit{(c) BJZZ 2016-2021}}  & \multicolumn{4}{c}{\textit{(d) QMP 2016-2021}} \\
     \cmidrule(lr){2-5} \cmidrule(lr){6-9}
     & \multicolumn{2}{c}{Intense Selling} & \multicolumn{2}{c}{Intense Buying}  & \multicolumn{2}{c}{Intense Selling} & \multicolumn{2}{c}{Intense Buying} \\ 
     \cmidrule(lr){2-3} \cmidrule(lr){4-5} \cmidrule(lr){6-7} \cmidrule(lr){8-9} 
         & Mean & \textit{t}-stat & Mean & \textit{t}-stat & Mean & \textit{t}-stat & Mean & \textit{t}-stat \\ 
  \midrule
        $k = -20$ & $-$0.0025$^{*}$ & $-$2.06 & $-$0.0179$^{**}$ & $-$12.93 & 0.0012 & 0.84 & $-$0.0219$^{**}$ & $-$14.46 \\ 
        $k = -15$ & $-$0.0009 & $-$0.90 & $-$0.0148$^{**}$ & $-$13.69 & 0.0020 & 1.65 & $-$0.0185$^{**}$ & $-$15.33 \\ 
        $k = -10$ & 0.0005 & 0.64 & $-$0.0114$^{**}$ & $-$14.88 & 0.0025$^{**}$ & 2.62 & $-$0.0145$^{**}$ & $-$15.72 \\ 
        $k = -5$ & 0.0009 & 1.80 & $-$0.0071$^{**}$ & $-$16.25 & 0.0022$^{**}$ & 4.07 & $-$0.0087$^{**}$ & $-$15.37 \\ 
        $k = 0$ & $-$0.0054$^{**}$ & $-$9.81 & $-$0.0003 & $-$0.60 & $-$0.0067$^{**}$ & $-$11.19 & 0.0028$^{**}$ & 5.41 \\ 
        $k = 5$ & $-$0.0013$^{**}$ & $-$3.16 & 0.0007 & 1.84 & $-$0.0020$^{**}$ & $-$4.46 & 0.0020$^{**}$ & 4.82 \\ 
        $k = 10$ & $-$0.0021$^{**}$ & $-$2.85 & 0.0014$^{*}$ & 2.17 & $-$0.0032$^{**}$ & $-$4.29 & 0.0023$^{**}$ & 3.76 \\ 
        $k = 15$ & $-$0.0032$^{**}$ & $-$3.34 & 0.0019$^{*}$ & 2.06 & $-$0.0046$^{**}$ & $-$4.51 & 0.0032$^{**}$ & 4.03 \\ 
        $k = 20$ & $-$0.0038$^{**}$ & $-$3.06 & 0.0020 & 1.90 & $-$0.0054$^{**}$ & $-$3.96 & 0.0033$^{**}$ & 3.51 \\ 
        \bottomrule
    \end{tabular}}
\end{table}
\restoregeometry


\newpage
\begin{center}
    \Large{
        \textbf{-- Online Appendix --}\\
        \textbf{``Revisiting \citet{Boehmeretal2021}: Recent Period, Alternative Method, Different Conclusions''}
        \\
        David Ardia, Clément Aymard, Tolga Cenesizoglu
    }
    \\ \vspace{1.5cm}
    \singlespacing
    \RaggedRight
    \normalsize
    This online appendix provides the following supplemental contents:
    \begin{itemize}
        \item Tables~\ref{tab:BJZZTAB1_REP}-\ref{tab:BJZZTAB8_REP}: Replication results of Tables I to VIII in BJZZ.
        \item Tables~\ref{tab:BJZZTAB2_COMPLETE}-\ref{tab:BJZZTAB8_COMPLETE}: Full set of results of Tables 2 to 8 presented in the main paper.
    \end{itemize}
\end{center}

\setcounter{section}{0}
\renewcommand*{\thesection}{\Roman{section}} 
\renewcommand*{\thesubsection}{\thesection.\Alph{subsection}} 
\setcounter{table}{0}
\renewcommand*{\thetable}{A\arabic{table}} 
\setcounter{figure}{0}
\renewcommand*{\thefigure}{A\arabic{figure}} 
\setcounter{page}{1}
\renewcommand*{\thepage}{\arabic{page}} 

\newpage

\newgeometry{left=2cm,right=2cm,top=2cm,bottom=2cm}

\begin{table}[!h]
\caption{\textbf{Summary statistics}}
\label{tab:BJZZTAB1_REP}
\centering
\scalebox{0.93}{
    \centering
\begin{tabular}{lcccccc}
    \toprule
    \multicolumn{7}{c}{Panel A: Original Results Reported in BJZZ} \\ 
         & N & Mean & Std & Median & Q1 & Q3 \\ \cmidrule(lr){1-7}
        \textsf{Mrbvol} & 4,628,957 & 42,481 & 280,474 & 5,165 & 1,200 & 20,681 \\ 
        \textsf{Mrsvol} & 4,628,957 & 42,430 & 264,704 & 5,635 & 1,369 & 21,828 \\ 
        \textsf{Mrbtrd} & 4,628,957 & 110 & 410 & 22 & 5 & 79 \\ 
        \textsf{Mrstrd} & 4,628,957 & 108 & 355 & 24 & 6 & 81 \\ 
        \textsf{Mroibvol} & 4,628,957 & -0.038 & 0.464 & -0.027 & -0.301 & 0.217 \\ 
        \textsf{Mroibtrd} & 4,628,957 & -0.032 & 0.437 & -0.010 & -0.276 & 0.205 \\ 
    \midrule
    \multicolumn{7}{c}{Panel B: Replication Results} \\ 
         & N & Mean & Std & Median & Q1 & Q3 \\ \cmidrule(lr){1-7}
        \textsf{Mrbvol} & 4,348,327 & 39,840 & 278,026 & 4,899 & 1,130 & 19,165 \\ 
        \textsf{Mrsvol} & 4,348,327 & 39,655 & 262,689 & 5,302 & 1,300 & 20,185 \\ 
        \textsf{Mrbtrd} & 4,348,327 & 99 & 386 & 21 & 5 & 72 \\ 
        \textsf{Mrstrd} & 4,348,327 & 97 & 330 & 22 & 6 & 74 \\ 
        \textsf{Mroibvol} & 4,348,327 & -0.036 & 0.470 & -0.025 & -0.304 & 0.224 \\ 
        \textsf{Mroibtrd} & 4,348,327 & -0.031 & 0.443 & -0.007 & -0.280 & 0.213 \\ 
        \bottomrule
    \end{tabular}}
\end{table}
\newpage

\begin{table}[!h]
\caption{\textbf{Determinants of Marketable Retail Order Imbalances}}
\label{tab:BJZZTAB2_REP}
\centering
\scalebox{0.9}{
    \centering
    \begin{tabular}{lrrrrrrrr}
    \toprule
     \multicolumn{9}{c}{Panel A: Original Results Reported in BJZZ} \\
     & \multicolumn{2}{c}{\textsf{Mroibvol}} 
     & \multicolumn{2}{c}{\textsf{Mroibvol}}
     & \multicolumn{2}{c}{\textsf{Mroibtrd}}
     & \multicolumn{2}{c}{\textsf{Mroibtrd}}\\ 
     & \multicolumn{2}{c}{Bid-Ask Return}
     & \multicolumn{2}{c}{CRSP Return}
     & \multicolumn{2}{c}{Bid-Ask Return}
     & \multicolumn{2}{c}{CRSP Return}\\
     \cmidrule(lr){2-3} \cmidrule(lr){4-5} \cmidrule(lr){6-7} \cmidrule(lr){8-9}
     & Coef. & \textit{t}-Stat & Coef. & \textit{t}-Stat & Coef. & \textit{t}-Stat & Coef. & \textit{t}-Stat \\ \midrule
        Intercept & $-$0.4013 & $-$20.03 & $-$0.4065 & $-$20.19 & $-$0.4326 & $-$22.00 & $-$0.4357 & $-$22.01 \\ 
        \textsf{Mroib}($w-1$) & 0.2200 & 92.53 & 0.2201 & 92.57 & 0.2865 & 150.01 & 0.2866 & 150.06 \\ 
        \textsf{Ret}($w-1$) & $-$0.9481 & $-$40.60 & $-$0.9620 & $-$41.43 & $-$0.9003 & $-$35.92 & $-$0.9156 & $-$36.74 \\ 
        \textsf{Ret}($m-1$) & $-$0.2778 & $-$19.24 & $-$0.2784 & $-$19.30 & $-$0.2258 & $-$14.84 & $-$0.2262 & $-$14.87 \\
\bottomrule
    \end{tabular}}
\end{table}
\newpage

\begin{table}[!h]
\caption{\textbf{Predicting Next-Week Returns Using Marketable Retail Order Imbalances}}
\label{tab:BJZZTAB3_REP}
\centering
\scalebox{0.9}{
    \centering
    \begin{tabular}{lrrrrrrrr}
   \toprule
     \multicolumn{9}{c}{Panel A: Original Results Reported in BJZZ} \\
     & \multicolumn{2}{c}{\textsf{Mroibvol}} 
     & \multicolumn{2}{c}{\textsf{Mroibvol}} 
     & \multicolumn{2}{c}{\textsf{Mroibtrd}}
     & \multicolumn{2}{c}{\textsf{Mroibtrd}}\\ 
     & \multicolumn{2}{c}{Bid-Ask Return} 
     & \multicolumn{2}{c}{CRSP Return}
     & \multicolumn{2}{c}{Bid-Ask Return}
     & \multicolumn{2}{c}{CRSP Return}\\
      \cmidrule(lr){2-3} \cmidrule(lr){4-5} \cmidrule(lr){6-7} \cmidrule(lr){8-9}
     & Coef. & \textit{t}-Stat & Coef. & \textit{t}-Stat & Coef. & \textit{t}-Stat & Coef. & \textit{t}-Stat \\ \midrule
        Intercept & 0.0050 & 2.58 & 0.0056 & 2.85 & 0.0050 & 2.58 & 0.0056 & 2.85 \\ 
        \textsf{Mroib}($w-1$) & 0.0009 & 15.60 & 0.0010 & 16.29 & 0.0008 & 12.30 & 0.0008 & 13.20 \\ 
        \textsf{Ret}($w-1$) & $-$0.0185 & $-$5.83 & $-$0.0220 & $-$6.85 & $-$0.0186 & $-$5.88 & $-$0.0222 & $-$6.91 \\ 
        \textsf{Ret}($m-1$) & 0.0006 & 0.35 & 0.0006 & 0.34 & 0.0005 & 0.29 & 0.0005 & 0.29 \\
\bottomrule
    \end{tabular}}
\end{table}
\newpage

\begin{table}[!h]
\caption{\textbf{Marketable Retail Return Predictability within Subgroups}
}
\label{tab:BJZZTAB4_REP}
\centering
\scalebox{0.9}{
    \centering
    \begin{tabular}{lrrrrrrrr}
   \toprule
    \multicolumn{9}{c}{Panel A: Original Results Reported in BJZZ} \\
    & \multicolumn{4}{c}{\textsf{Mroibvol}} & \multicolumn{4}{c}{\textsf{Mroibtrd}} \\
    \cmidrule(lr){2-5} \cmidrule(lr){6-9}
         & Coef. & \textit{t}-Stat & IQR &  W.R. Diff. & Coef. & \textit{t}-Stat & IQR & W.R. Diff. \\ 
        \midrule
        \multicolumn{9}{l}{\textit{Market-cap subgroups}} \\
        Small & 0.0013 & 13.90 & 1.6630 & 0.219\% & 0.0012 & 11.58 & 1.7360 & 0.207\% \\ 
        Medium & 0.0007 & 9.18 & 1.3230 & 0.087\% & 0.0004 & 5.63 & 1.3460 & 0.059\% \\ 
        Big & 0.0003 & 3.68 & 0.8920 & 0.026\% & 0.0002 & 2.52 & 0.9290 & 0.019\% \\ 
        \multicolumn{9}{l}{\textit{Share-price subgroups}} \\
\bottomrule
    \end{tabular}}
\end{table}%
\newpage

\begin{table}[!h]
\caption{\textbf{Predicting Returns k Weeks Ahead}}
\label{tab:BJZZTAB5_REP}
\centering
\scalebox{0.9}{
    \centering
    \begin{tabular}{lcccc}
    \toprule
    \multicolumn{5}{c}{Panel A: Original Results Reported in BJZZ} \\
     & \multicolumn{2}{c}{\textsf{Mroibvol}} & \multicolumn{2}{c}{\textsf{Mroibtrd}}\\
    \cmidrule(lr){2-3} \cmidrule(lr){4-5}
         & Coef. & \textit{t}-Stat & Coef. & \textit{t}-Stat \\ 
        \midrule
        \multicolumn{5}{l}{\textit{Bid-ask average returns}} \\
        1 week & 0.0009 & 15.60 & 0.0008 & 12.30 \\ 
        2 weeks & 0.0006 & 9.35 & 0.0005 & 7.89 \\ 
        4 weeks & 0.0003 & 5.56 & 0.0003 & 4.66 \\ 
        6 weeks & 0.0002 & 3.90 & 0.0002 & 2.60 \\
\bottomrule
    \end{tabular}}
\end{table}
\newpage

\begin{table}[!h]
\caption{\textbf{Long-Short Strategy Returns Based on Marketable Retail Order Imbalances}}
\label{tab:BJZZTAB6_REP}
\centering
\scalebox{0.9}{
    \centering
    \begin{tabular}{lrrrrrrrrrr}
    \toprule
    \multicolumn{11}{c}{Panel A: Original Results Reported in BJZZ} \\
     & \multicolumn{4}{c}{All Stocks} & \multicolumn{2}{c}{Small}
     & \multicolumn{2}{c}{Medium} & \multicolumn{2}{c}{Big} \\ 
     \cmidrule(lr){2-5} \cmidrule(lr){6-7} \cmidrule(lr){8-9} \cmidrule(lr){10-11}
         & Mean & \textit{t}-Stat & Alpha & \textit{t}-Stat & Alpha & \textit{t}-Stat & Alpha & \textit{t}-Stat & Alpha & \textit{t}-Stat \\ 
        \midrule
        \multicolumn{11}{l}{\textit{Form Portfolios on the Previous Week’s \textsf{Mroibvol}}} \\ 
        1 week & 0.092\% & 2.66 & 0.084\% & 2.43 & 0.403\% & 9.16 & 0.170\% & 6.24 & 0.067\% & 1.78 \\ 
        2 weeks & 0.147\% & 2.45 & 0.135\% & 2.46 & 0.669\% & 9.01 & 0.292\% & 6.81 & 0.105\% & 1.70 \\ 
        4 weeks & 0.223\% & 1.89 & 0.208\% & 2.00 & 1.124\% & 10.43 & 0.423\% & 6.36 & 0.143\% & 1.22 \\ 
        6 weeks & 0.310\% & 1.72 & 0.277\% & 1.73 & 1.399\% & 13.02 & 0.558\% & 6.07 & 0.171\% & 1.05 \\
\bottomrule
    \end{tabular}}
\end{table}
\newpage

\begin{landscape}
\begin{table}[!h]
\caption{\textbf{Predictability Decomposition}}
\label{tab:BJZZTAB7_REP}
\centering
\scalebox{0.9}{
    \centering
    \begin{tabular}{lrrrrrrrr}
    \toprule
    \multicolumn{9}{c}{Panel A: Original Results Reported in BJZZ} \\
    \\
    \multicolumn{9}{l}{\textit{First Stage of Projecting Order Imbalance on Persistence and Past Returns}} \\ 
    & \multicolumn{2}{c}{$\textsf{Mroivol}(w-1)$} & \multicolumn{2}{c}{$\textsf{Mroivol}(w-1)$} 
    & \multicolumn{2}{c}{$\textsf{Mroitrd}(w-1)$} & \multicolumn{2}{c}{$\textsf{Mroitrd}(w-1)$} \\ 
     & \multicolumn{2}{c}{Bid-Ask Return} & \multicolumn{2}{c}{CRSP Return}
     & \multicolumn{2}{c}{Bid-Ask Return} & \multicolumn{2}{c}{CRSP Return} \\
     \cmidrule(lr){2-3} \cmidrule(lr){4-5} \cmidrule(lr){6-7} \cmidrule(lr){8-9} 
    & Coef. & \textit{t}-Stat & Coef. & \textit{t}-Stat & Coef. & \textit{t}-Stat & Coef. & \textit{t}-Stat \\ 
        \midrule
        Intercept & $-$0.1413 & $-$24.66 & $-$0.1408 & $-$24.61 & $-$0.1054 & $-$17.23 & $-$0.1049 & $-$17.19 \\ 
        $\textsf{Mroivol}(w-2)$ & 0.2227 & 96.20 & 0.2228 & 96.20 & 0.2906 & 149.82 & 0.2907 & 149.85 \\
\end{tabular}}
\end{table}%
\newpage
\begin{table}[!h]\ContinuedFloat
\caption{\textbf{Predictability Decomposition (Continued)}}
\centering
\scalebox{0.9}{
    \centering
    \begin{tabular}{lrrrrrrrr}
    \midrule
    \multicolumn{9}{c}{Panel B: Replication Results} \\
    \\
    \multicolumn{9}{l}{\textit{First Stage of Projecting Order Imbalance on Persistence and Past Returns}} \\ 
    & \multicolumn{2}{c}{$\textsf{Mroivol}(w-1)$} & \multicolumn{2}{c}{$\textsf{Mroivol}(w-1)$} 
    & \multicolumn{2}{c}{$\textsf{Mroitrd}(w-1)$} & \multicolumn{2}{c}{$\textsf{Mroitrd}(w-1)$} \\ 
     & \multicolumn{2}{c}{Bid-Ask Return} & \multicolumn{2}{c}{CRSP Return}
     & \multicolumn{2}{c}{Bid-Ask Return} & \multicolumn{2}{c}{CRSP Return} \\
     \cmidrule(lr){2-3} \cmidrule(lr){4-5} \cmidrule(lr){6-7} \cmidrule(lr){8-9} 
    & Coef. & \textit{t}-Stat & Coef. & \textit{t}-Stat & Coef. & \textit{t}-Stat & Coef. & \textit{t}-Stat \\ 
        \midrule
        Intercept & $-$0.1351 & $-$23.28 & $-$0.1350 & $-$23.27 & $-$0.1127 & $-$12.63 & $-$0.1126 & $-$12.62 \\ 
        $\textsf{Mroivol}(w-2)$ & 0.2014 & 78.12 & 0.2015 & 78.13 & 0.2743 & 97.13 & 0.2744 & 97.13 \\
\bottomrule
    \end{tabular}}
\end{table}
\end{landscape}
\newpage

\begin{table}[!h]
\caption{\textbf{Marketable Retail Order Imbalance and Contemporaneous Returns}}
\label{tab:BJZZTAB8_REP}
\centering
\scalebox{0.9}{
    \centering
    \begin{tabular}{lrrrrrrrr}
    \toprule
    \multicolumn{9}{c}{Panel A: Original Results Reported in BJZZ} \\
     & \multicolumn{2}{c}{Intense Selling} & \multicolumn{2}{c}{Selling}
     & \multicolumn{2}{c}{Buying} & \multicolumn{2}{c}{Intense Buying} \\ 
     \cmidrule(lr){2-3} \cmidrule(lr){4-5} \cmidrule(lr){6-7} \cmidrule(lr){8-9} 
         & Mean & \textit{t}-Stat & Mean & \textit{t}-Stat & Mean & \textit{t}-Stat & Mean & \textit{t}-Stat \\ 
        \midrule
        \multicolumn{9}{l}{\textit{Cumulative Market-Adjusted Return}} \\ 
        $k = -20$ & 0.0067$^{**}$ & 5.48 & 0.0063$^{**}$ & 8.10 & $-$0.0111$^{**}$ & $-$19.00 & $-$0.0129$^{**}$ & $-$12.49 \\ 
        $k = -15$ & 0.0056$^{**}$ & 5.62 & 0.0055$^{**}$ & 8.71 & $-$0.0096$^{**}$ & $-$20.87 & $-$0.0109$^{**}$ & $-$12.91 \\ 
        $k = -10$ & 0.0041$^{**}$ & 5.40 & 0.0042$^{**}$ & 9.04 & $-$0.0074$^{**}$ & $-$20.83 & $-$0.0084$^{**}$ & $-$12.93 \\ 
        $k = -5$ & 0.0027$^{**}$ & 6.06 & 0.0028$^{**}$ & 10.07 & $-$0.0047$^{**}$ & $-$24.09 & $-$0.0053$^{**}$ & $-$15.62 \\
\bottomrule
    \end{tabular}}
\end{table}

\restoregeometry
\newpage

\newgeometry{left=1.5cm,right=1.5cm,top=2cm,bottom=2cm}

\begin{table}[!h]
\caption{\textbf{Determinants of Marketable Retail Order Imbalances}}
\label{tab:BJZZTAB2_COMPLETE}
\centering
\scalebox{0.93}{
    \centering
\begin{tabular}{lrrrrrrrr}
    \toprule
     \multicolumn{9}{c}{\textit{(a) BJZZ 2010-2015}} \\
     & \multicolumn{2}{c}{\textsf{Mroibvol}} 
     & \multicolumn{2}{c}{\textsf{Mroibvol}}
     & \multicolumn{2}{c}{\textsf{Mroibtrd}}
     & \multicolumn{2}{c}{\textsf{Mroibtrd}}\\ 
     & \multicolumn{2}{c}{Bid-Ask Return}
     & \multicolumn{2}{c}{CRSP Return}
     & \multicolumn{2}{c}{Bid-Ask Return}
     & \multicolumn{2}{c}{CRSP Return}\\
     \cmidrule(lr){2-3} \cmidrule(lr){4-5} \cmidrule(lr){6-7} \cmidrule(lr){8-9}
     & Coef. & \textit{t}-Stat & Coef. & \textit{t}-Stat & Coef. & \textit{t}-Stat & Coef. & \textit{t}-Stat \\ \midrule
        Intercept & $-$0.2833 & $-$22.23 & $-$0.2848 & $-$22.13 & $-$0.2866 & $-$21.02 & $-$0.2868 & $-$20.91 \\ 
        \textsf{Mroib}($w-1$) & 0.1982 & 71.81 & 0.1982 & 71.82 & 0.2698 & 91.06 & 0.2698 & 91.07 \\ 
        \textsf{Ret}($w-1$) & $-$0.8302 & $-$35.91 & $-$0.8311 & $-$36.26 & $-$0.7782 & $-$31.41 & $-$0.7811 & $-$31.85 \\ 
        \textsf{Ret}($m-1$) & $-$0.1680 & $-$13.08 & $-$0.1681 & $-$13.08 & $-$0.1214 & $-$8.91 & $-$0.1213 & $-$8.91 \\
\end{tabular}}
\end{table}%
\newpage
\begin{table}[!h]\ContinuedFloat
\caption{\textbf{Determinants of Marketable Retail Order Imbalances (Continued)}}
\centering
\scalebox{0.93}{
    \centering
    \begin{tabular}{lrrrrrrrr}
    \midrule
     \multicolumn{9}{c}{\textit{(c) BJZZ 2016-2021}} \\
     & \multicolumn{2}{c}{\textsf{Mroibvol}} 
     & \multicolumn{2}{c}{\textsf{Mroibvol}}
     & \multicolumn{2}{c}{\textsf{Mroibtrd}}
     & \multicolumn{2}{c}{\textsf{Mroibtrd}}\\ 
     & \multicolumn{2}{c}{Bid-Ask Return}
     & \multicolumn{2}{c}{CRSP Return}
     & \multicolumn{2}{c}{Bid-Ask Return}
     & \multicolumn{2}{c}{CRSP Return}\\
     \cmidrule(lr){2-3} \cmidrule(lr){4-5} \cmidrule(lr){6-7} \cmidrule(lr){8-9}
     & Coef. & \textit{t}-Stat & Coef. & \textit{t}-Stat & Coef. & \textit{t}-Stat & Coef. & \textit{t}-Stat \\
     \midrule
        Intercept & $-$0.2072 & $-$27.82 & $-$0.2075 & $-$27.68 & $-$0.1472 & $-$19.54 & $-$0.1468 & $-$19.54 \\ 
        \textsf{Mroib}($w-1$) & 0.0983 & 57.30 & 0.0984 & 57.30 & 0.2237 & 84.06 & 0.2238 & 84.08 \\ 
        \textsf{Ret}($w-1$) & $-$0.3567 & $-$25.85 & $-$0.3538 & $-$25.83 & $-$0.4100 & $-$28.71 & $-$0.4076 & $-$28.66 \\ 
         \textsf{Ret}($m-1$) & $-$0.0933 & $-$12.49 & $-$0.0928 & $-$12.40 & $-$0.1023 & $-$13.87 & $-$0.1016 & $-$13.72 \\
\bottomrule
    \end{tabular}}
\end{table}
\newpage 

\begin{table}[!h]
\caption{\textbf{Predicting Next-Week Returns Using Marketable Retail Order Imbalances}}
\label{tab:BJZZTAB3_COMPLETE}
\centering
\scalebox{0.93}{
    \centering
    \begin{tabular}{lrrrrrrrr}
    \toprule
     & \multicolumn{8}{c}{\textit{(a) BJZZ 2010-2015}} \\
          & \multicolumn{2}{c}{\textsf{Mroibvol}} 
     & \multicolumn{2}{c}{\textsf{Mroibvol}} 
     & \multicolumn{2}{c}{\textsf{Mroibtrd}}
     & \multicolumn{2}{c}{\textsf{Mroibtrd}}\\ 
     & \multicolumn{2}{c}{Bid-Ask Return} 
     & \multicolumn{2}{c}{CRSP Return}
     & \multicolumn{2}{c}{Bid-Ask Return}
     & \multicolumn{2}{c}{CRSP Return}\\
      \cmidrule(lr){2-3} \cmidrule(lr){4-5} \cmidrule(lr){6-7} \cmidrule(lr){8-9}
     & Coef. & \textit{t}-Stat & Coef. & \textit{t}-Stat & Coef. & \textit{t}-Stat & Coef. & \textit{t}-Stat \\ \midrule
        Intercept & 0.0033 & 2.24 & 0.0036 & 2.42 & 0.0033 & 2.23 & 0.0036 & 2.40 \\ 
        \textsf{Mroib}($w-1$) & 0.0009 & 15.14 & 0.0009 & 14.55 & 0.0008 & 11.93 & 0.0008 & 11.41 \\ 
        \textsf{Ret}($w-1$) & $-$0.0172 & $-$5.45 & $-$0.0206 & $-$6.50 & $-$0.0174 & $-$5.50 & $-$0.0208 & $-$6.56 \\ 
        \textsf{Ret}($m-1$) & $-$0.0001 & $-$0.03 & $-$0.0003 & $-$0.16 & $-$0.0001 & $-$0.07 & $-$0.0004 & $-$0.20 \\
\end{tabular}}
\end{table}%
\newpage
\begin{table}[!h]\ContinuedFloat
\caption{\textbf{Predicting Next-Week Returns Using Marketable Retail Order Imbalances (Continued)}}
\centering
\scalebox{0.93}{
    \centering
    \begin{tabular}{lrrrrrrrr}
    \midrule
     & \multicolumn{8}{c}{\textit{(c) BJZZ 2016-2021}} \\
          & \multicolumn{2}{c}{\textsf{Mroibvol}} 
     & \multicolumn{2}{c}{\textsf{Mroibvol}} 
     & \multicolumn{2}{c}{\textsf{Mroibtrd}}
     & \multicolumn{2}{c}{\textsf{Mroibtrd}}\\ 
     & \multicolumn{2}{c}{Bid-Ask Return} 
     & \multicolumn{2}{c}{CRSP Return}
     & \multicolumn{2}{c}{Bid-Ask Return}
     & \multicolumn{2}{c}{CRSP Return}\\
      \cmidrule(lr){2-3} \cmidrule(lr){4-5} \cmidrule(lr){6-7} \cmidrule(lr){8-9}
     & Coef. & \textit{t}-Stat & Coef. & \textit{t}-Stat & Coef. & \textit{t}-Stat & Coef. & \textit{t}-Stat \\ \midrule
        Intercept & 0.0041 & 2.24 & 0.0045 & 2.42 & 0.0042 & 2.25 & 0.0045 & 2.43 \\ 
        \textsf{Mroib}($w-1$) & 0.0006 & 7.95 & 0.0006 & 7.54 & 0.0007 & 4.86 & 0.0006 & 4.49 \\ 
        \textsf{Ret}($w-1$) & $-$0.0138 & $-$3.73 & $-$0.0174 & $-$4.74 & $-$0.0138 & $-$3.74 & $-$0.0174 & $-$4.74 \\ 
        \textsf{Ret}($m-1$) & $-$0.0019 & $-$0.98 & $-$0.0020 & $-$1.03 & $-$0.0019 & $-$0.97 & $-$0.0020 & $-$1.02 \\
\bottomrule
    \end{tabular}}
\end{table}
\newpage 

\begin{table}[!h]
\caption{\textbf{Marketable Retail Return Predictability within Subgroups}}
\label{tab:BJZZTAB4_COMPLETE}
\centering
\scalebox{0.93}{
    \centering
    \begin{tabular}{lcccccccc}
    \toprule
     & \multicolumn{8}{c}{\textit{(a) BJZZ 2010-2015}} \\
    & \multicolumn{4}{c}{\textsf{Mroibvol}} & \multicolumn{4}{c}{\textsf{Mroibtrd}} \\
    \cmidrule(lr){2-5} \cmidrule(lr){6-9}
         & Coef. & \textit{t}-Stat & IQR &  W.R. Diff. & Coef. & \textit{t}-Stat & IQR & W.R. Diff. \\ 
        \midrule
        \multicolumn{9}{l}{\textit{Market-cap subgroups}} \\
        Small & 0.0013 & 13.87 & 1.6010 & 0.205\% & 0.0011 & 11.12 & 1.6768 & 0.184\% \\ 
        Medium & 0.0005 & 6.70 & 1.2386 & 0.068\% & 0.0004 & 4.89 & 1.2411 & 0.054\% \\ 
        Big & 0.0003 & 3.79 & 0.8746 & 0.028\% & 0.0003 & 2.85 & 0.8972 & 0.023\% \\ 
        \multicolumn{9}{l}{\textit{Share-price subgroups}} \\
\end{tabular}}
\end{table}%
\newpage
\begin{table}[!h]\ContinuedFloat
\caption{\textbf{Marketable Retail Return Predictability within Subgroups (Continued)}}
\centering
\scalebox{0.93}{
    \centering
    \begin{tabular}{lcccccccc}
    \midrule
     & \multicolumn{8}{c}{\textit{(c) BJZZ 2016-2021}} \\
    & \multicolumn{4}{c}{\textsf{Mroibvol}} & \multicolumn{4}{c}{\textsf{Mroibtrd}} \\
    \cmidrule(lr){2-5} \cmidrule(lr){6-9}
         & Coef. & \textit{t}-Stat & IQR &  W.R. Diff. & Coef. & \textit{t}-Stat & IQR & W.R. Diff. \\ 
        \midrule
        \multicolumn{9}{l}{\textit{Market-cap subgroups}} \\
        Small & 0.0007 & 6.4400 & 1.3347 & 0.098\% & 0.0008 & 4.6500 & 1.1714 & 0.094\% \\ 
        Medium & 0.0004 & 4.0300 & 1.0890 & 0.045\% & 0.0006 & 3.4800 & 0.8699 & 0.053\% \\ 
        Big & 0.0002 & 1.5400 & 0.6799 & 0.014\% & -0.0001 & -0.6600 & 0.5317 & -0.006\% \\
        \multicolumn{9}{l}{\textit{Share-price subgroups}} \\
\bottomrule
    \end{tabular}}
\end{table}%
\newpage 

\begin{table}[!h]
\caption{\textbf{Predicting Returns k Weeks Ahead}}
\label{tab:BJZZTAB5_COMPLETE}
\centering
\scalebox{0.9}{
    \centering
    \begin{tabular}{lcccccccc}
    \toprule
    & \multicolumn{4}{c}{\textit{(a) BJZZ 2010-2015}} & \multicolumn{4}{c}{\textit{(b) QMP 2010-2015}} \\
    & \multicolumn{2}{c}{\textsf{Mroibvol}} & \multicolumn{2}{c}{\textsf{Mroibtrd}} & \multicolumn{2}{c}{\textsf{Mroibvol}} & \multicolumn{2}{c}{\textsf{Mroibtrd}}\\
    \cmidrule(lr){2-3} \cmidrule(lr){4-5} \cmidrule(lr){6-7} \cmidrule(lr){8-9}
    & Coef. & \textit{t}-Stat & Coef. & \textit{t}-Stat & Coef. & \textit{t}-Stat & Coef. & \textit{t}-Stat \\ 
        \midrule
        \multicolumn{5}{l}{\textit{Bid-ask average returns}} \\
        1 week & 0.0009 & 15.14 & 0.0008 & 11.93 & 0.0010 & 14.84 & 0.0009 & 12.13 \\ 
        2 weeks & 0.0006 & 9.48 & 0.0005 & 7.77 & 0.0006 & 9.00 & 0.0005 & 7.09 \\ 
        4 weeks & 0.0003 & 5.64 & 0.0003 & 5.40 & 0.0003 & 5.52 & 0.0003 & 5.59 \\ 
        6 weeks & 0.0003 & 4.53 & 0.0002 & 3.31 & 0.0003 & 4.64 & 0.0002 & 3.67 \\
\bottomrule
    \end{tabular}}
\end{table}
\newpage 

\begin{table}[!h]
\caption{\textbf{Long-Short Strategy Returns Based on Marketable Retail Order Imbalances}}
\label{tab:BJZZTAB6_COMPLETE}
\centering
\scalebox{0.9}{
    \centering
    \begin{tabular}{lrrrrrrrrrr}
    \toprule
    & \multicolumn{10}{c}{\textit{(a) BJZZ 2010-2015}} \\
     & \multicolumn{4}{c}{All Stocks} & \multicolumn{2}{c}{Small}
     & \multicolumn{2}{c}{Medium} & \multicolumn{2}{c}{Big} \\ 
     \cmidrule(lr){2-5} \cmidrule(lr){6-7} \cmidrule(lr){8-9} \cmidrule(lr){10-11}
         & Mean & \textit{t}-Stat & Alpha & \textit{t}-Stat & Alpha & \textit{t}-Stat & Alpha & \textit{t}-Stat & Alpha & \textit{t}-Stat \\ 
        \midrule
        \multicolumn{11}{l}{\textit{Form Portfolios on the Previous Week’s \textsf{Mroibvol}}} \\ 
        1 week & 0.095\% & 3.27 & 0.083\% & 2.77 & 0.437\% & 10.39 & 0.175\% & 5.59 & 0.051\% & 1.52 \\ 
        2 weeks & 0.124\% & 2.38 & 0.090\% & 1.81 & 0.613\% & 8.68 & 0.270\% & 5.01 & 0.052\% & 1.04 \\ 
        4 weeks & 0.226\% & 2.67 & 0.167\% & 2.04 & 0.852\% & 7.15 & 0.377\% & 4.46 & 0.125\% & 1.54 \\ 
        6 weeks & 0.347\% & 2.84 & 0.285\% & 2.56 & 0.909\% & 6.54 & 0.471\% & 3.88 & 0.193\% & 1.76 \\
\end{tabular}}
\end{table}%
\newpage
\begin{table}[!h]\ContinuedFloat
\caption{\textbf{Long-Short Strategy Returns Based on Marketable Retail Order Imbalances (Continued)}}
\centering
\scalebox{0.9}{
    \centering
    \begin{tabular}{lrrrrrrrrrr}
    \midrule
    & \multicolumn{10}{c}{\textit{(c) BJZZ 2016-2021}} \\
     & \multicolumn{4}{c}{All Stocks} & \multicolumn{2}{c}{Small}
     & \multicolumn{2}{c}{Medium} & \multicolumn{2}{c}{Big} \\ 
     \cmidrule(lr){2-5} \cmidrule(lr){6-7} \cmidrule(lr){8-9} \cmidrule(lr){10-11}
         & Mean & \textit{t}-Stat & Alpha & \textit{t}-Stat & Alpha & \textit{t}-Stat & Alpha & \textit{t}-Stat & Alpha & \textit{t}-Stat \\ 
        \midrule
        \multicolumn{11}{l}{\textit{Form Portfolios on the Previous Week’s \textsf{Mroibvol}}} \\ 
        1 week & $-$0.008\% & $-$0.18 & $-$0.021\% & $-$0.48 & 0.143\% & 3.16 & 0.067\% & 1.61 & $-$0.042\% & $-$0.91 \\ 
        2 weeks & 0.063\% & 1.04 & 0.004\% & 0.07 & 0.177\% & 2.54 & 0.025\% & 0.41 & $-$0.075\% & $-$1.10 \\ 
        4 weeks & 0.105\% & 0.99 & $-$0.018\% & $-$0.20 & 0.381\% & 2.75 & $-$0.013\% & $-$0.14 & $-$0.114\% & $-$1.01 \\ 
        6 weeks & 0.086\% & 0.58 & $-$0.126\% & $-$1.02 & 0.424\% & 2.56 & $-$0.128\% & $-$1.24 & $-$0.211\% & $-$1.62 \\
\bottomrule
    \end{tabular}}
\end{table}
\newpage 

\begin{landscape}
\begin{table}[!h]
\caption{\textbf{Predictability Decomposition}}
\label{tab:BJZZTAB7_REP}
\centering
\scalebox{0.9}{
    \centering
    \begin{tabular}{lrrrrrrrr}
    \toprule
    \multicolumn{9}{c}{\textit{(a) BJZZ 2010-2015}} \\
    \\
\multicolumn{9}{l}{\textit{First Stage of Projecting Order Imbalance on Persistence and Past Returns}} \\ 
    & \multicolumn{2}{c}{$\textsf{Mroivol}(w-1)$} & \multicolumn{2}{c}{$\textsf{Mroivol}(w-1)$} 
    & \multicolumn{2}{c}{$\textsf{Mroitrd}(w-1)$} & \multicolumn{2}{c}{$\textsf{Mroitrd}(w-1)$} \\ 
     & \multicolumn{2}{c}{Bid-Ask Return} & \multicolumn{2}{c}{CRSP Return}
     & \multicolumn{2}{c}{Bid-Ask Return} & \multicolumn{2}{c}{CRSP Return} \\
     \cmidrule(lr){2-3} \cmidrule(lr){4-5} \cmidrule(lr){6-7} \cmidrule(lr){8-9} 
    & Coef. & \textit{t}-Stat & Coef. & \textit{t}-Stat & Coef. & \textit{t}-Stat & Coef. & \textit{t}-Stat \\ 
        \midrule
        Intercept & $-$0.1351 & $-$23.28 & $-$0.1350 & $-$23.27 & $-$0.1127 & $-$12.63 & $-$0.1126 & $-$12.62 \\ 
        $\textsf{Mroivol}(w-2)$ & 0.2014 & 78.12 & 0.2015 & 78.13 & 0.2743 & 97.13 & 0.2744 & 97.13 \\
\end{tabular}}
\end{table}%
\begin{table}[!h]\ContinuedFloat
\caption{\textbf{Predictability Decomposition (Continued)}}
\centering
\scalebox{0.9}{
    \centering
    \begin{tabular}{lrrrrrrrr}
    \midrule
    \multicolumn{9}{c}{\textit{(b) QMP 2010-2015}} \\
    \\
    \multicolumn{9}{l}{\textit{First Stage of Projecting Order Imbalance on Persistence and Past Returns}} \\ 
    & \multicolumn{2}{c}{$\textsf{Mroivol}(w-1)$} & \multicolumn{2}{c}{$\textsf{Mroivol}(w-1)$} 
    & \multicolumn{2}{c}{$\textsf{Mroitrd}(w-1)$} & \multicolumn{2}{c}{$\textsf{Mroitrd}(w-1)$} \\ 
     & \multicolumn{2}{c}{Bid-Ask Return} & \multicolumn{2}{c}{CRSP Return}
     & \multicolumn{2}{c}{Bid-Ask Return} & \multicolumn{2}{c}{CRSP Return} \\
     \cmidrule(lr){2-3} \cmidrule(lr){4-5} \cmidrule(lr){6-7} \cmidrule(lr){8-9} 
    & Coef. & \textit{t}-Stat & Coef. & \textit{t}-Stat & Coef. & \textit{t}-Stat & Coef. & \textit{t}-Stat \\ 
        \midrule
        Intercept & $-$0.1507 & $-$21.98 & $-$0.1505 & $-$21.97 & $-$0.1094 & $-$13.80 & $-$0.1092 & $-$13.79 \\ 
        $\textsf{Mroibvol}(w-2)$ & 0.2417 & 91.63 & 0.2418 & 91.64 & 0.2945 & 98.54 & 0.2946 & 98.53 \\
\end{tabular}}
\end{table}%
\begin{table}[!h]\ContinuedFloat
\caption{\textbf{Predictability Decomposition (Continued)}}
\centering
\scalebox{0.9}{
    \centering
    \begin{tabular}{lrrrrrrrr}
    \midrule
    \multicolumn{9}{c}{\textit{(c) BJZZ 2016-2021}} \\
    \\
    \multicolumn{9}{l}{\textit{First Stage of Projecting Order Imbalance on Persistence and Past Returns}} \\ 
    & \multicolumn{2}{c}{$\textsf{Mroivol}(w-1)$} & \multicolumn{2}{c}{$\textsf{Mroivol}(w-1)$} 
    & \multicolumn{2}{c}{$\textsf{Mroitrd}(w-1)$} & \multicolumn{2}{c}{$\textsf{Mroitrd}(w-1)$} \\ 
     & \multicolumn{2}{c}{Bid-Ask Return} & \multicolumn{2}{c}{CRSP Return}
     & \multicolumn{2}{c}{Bid-Ask Return} & \multicolumn{2}{c}{CRSP Return} \\
     \cmidrule(lr){2-3} \cmidrule(lr){4-5} \cmidrule(lr){6-7} \cmidrule(lr){8-9} 
    & Coef. & \textit{t}-Stat & Coef. & \textit{t}-Stat & Coef. & \textit{t}-Stat & Coef. & \textit{t}-Stat \\ 
        \midrule
        Intercept & $-$0.0804 & $-$22.69 & $-$0.0803 & $-$22.70 & $-$0.0209 & $-$4.70 & $-$0.0208 & $-$4.68 \\ 
        $\textsf{Mroibvol}(w-2)$ & 0.1003 & 60.92 & 0.1003 & 60.92 & 0.2281 & 89.64 & 0.2281 & 89.63 \\
\end{tabular}}
\end{table}%
\begin{table}[!h]\ContinuedFloat
\caption{\textbf{Predictability Decomposition (Continued)}}
\centering
\scalebox{0.9}{
    \centering
    \begin{tabular}{lrrrrrrrr}
    \midrule
    \multicolumn{9}{c}{\textit{(d) QMP 2016-2021}} \\
    \\
    \multicolumn{9}{l}{\textit{First Stage of Projecting Order Imbalance on Persistence and Past Returns}} \\ 
    & \multicolumn{2}{c}{$\textsf{Mroivol}(w-1)$} & \multicolumn{2}{c}{$\textsf{Mroivol}(w-1)$} 
    & \multicolumn{2}{c}{$\textsf{Mroitrd}(w-1)$} & \multicolumn{2}{c}{$\textsf{Mroitrd}(w-1)$} \\ 
     & \multicolumn{2}{c}{Bid-Ask Return} & \multicolumn{2}{c}{CRSP Return}
     & \multicolumn{2}{c}{Bid-Ask Return} & \multicolumn{2}{c}{CRSP Return} \\
     \cmidrule(lr){2-3} \cmidrule(lr){4-5} \cmidrule(lr){6-7} \cmidrule(lr){8-9} 
    & Coef. & \textit{t}-Stat & Coef. & \textit{t}-Stat & Coef. & \textit{t}-Stat & Coef. & \textit{t}-Stat \\ 
        \midrule
        Intercept & $-$0.1436 & $-$24.95 & $-$0.1435 & $-$24.94 & $-$0.0268 & $-$3.81 & $-$0.0266 & $-$3.79 \\ 
        $\textsf{Mroibvol}(w-2)$ & 0.1794 & 86.84 & 0.1794 & 86.84 & 0.3061 & 138.00 & 0.3062 & 137.97 \\
\bottomrule
    \end{tabular}}
\end{table}
\end{landscape}
\newpage 

\begin{table}[!h]
\caption{\textbf{Marketable Retail Order Imbalance and Contemporaneous Returns}}
\label{tab:BJZZTAB8_COMPLETE}
\centering
\scalebox{0.9}{
    \centering
    \begin{tabular}{lrrrrrrrr}
    \toprule
    \multicolumn{9}{c}{\textit{(a) BJZZ 2010-2015}} \\
    & \multicolumn{2}{c}{Intense Selling} & \multicolumn{2}{c}{Selling}
     & \multicolumn{2}{c}{Buying} & \multicolumn{2}{c}{Intense Buying} \\ 
     \cmidrule(lr){2-3} \cmidrule(lr){4-5} \cmidrule(lr){6-7} \cmidrule(lr){8-9} 
         & Mean & \textit{t}-Stat & Mean & \textit{t}-Stat & Mean & \textit{t}-Stat & Mean & \textit{t}-Stat \\ 
        \midrule
        \multicolumn{9}{l}{\textit{Cumulative Market-Adjusted Return}} \\ 
        $k = -20$ & 0.0074$^{**}$ & 7.33 & 0.0066$^{**}$ & 8.18 & $-$0.0127$^{**}$ & $-$15.39 & $-$0.0166$^{**}$ & $-$19.30 \\ 
        $k = -15$ & 0.0071$^{**}$ & 9.34 & 0.0064$^{**}$ & 10.94 & $-$0.0108$^{**}$ & $-$17.91 & $-$0.0137$^{**}$ & $-$21.05 \\ 
        $k = -10$ & 0.0059$^{**}$ & 10.89 & 0.0055$^{**}$ & 13.04 & $-$0.0083$^{**}$ & $-$19.19 & $-$0.0103$^{**}$ & $-$20.67 \\ 
        $k = -5$ & 0.0039$^{**}$ & 12.56 & 0.0035$^{**}$ & 15.59 & $-$0.0051$^{**}$ & $-$19.37 & $-$0.0064$^{**}$ & $-$20.74 \\
\end{tabular}}
\end{table}%
\newpage
\begin{table}[!h]\ContinuedFloat
\caption{\textbf{Marketable Retail Order Imbalance and Contemporaneous Returns (Continued)}}
\centering
\scalebox{0.9}{
    \centering
    \begin{tabular}{lrrrrrrrr}
    \midrule
\multicolumn{9}{c}{\textit{(c) BJZZ 2016-2021}} \\
     & \multicolumn{2}{c}{Intense Selling} & \multicolumn{2}{c}{Selling}
     & \multicolumn{2}{c}{Buying} & \multicolumn{2}{c}{Intense Buying} \\ 
     \cmidrule(lr){2-3} \cmidrule(lr){4-5} \cmidrule(lr){6-7} \cmidrule(lr){8-9} 
         & Mean & \textit{t}-Stat & Mean & \textit{t}-Stat & Mean & \textit{t}-Stat & Mean & \textit{t}-Stat \\ 
        \midrule
        \multicolumn{9}{l}{\textit{Cumulative Market-Adjusted Return}} \\ 
        $k = -20$ & $-$0.0025$^{*}$ & $-$2.06 & $-$0.0008 & $-$1.04 & $-$0.0142$^{**}$ & $-$14.76 & $-$0.0179$^{**}$ & $-$12.93 \\ 
        $k = -15$ & $-$0.0009 & $-$0.90 & 0.0002 & 0.36 & $-$0.0117$^{**}$ & $-$16.10 & $-$0.0148$^{**}$ & $-$13.69 \\ 
        $k = -10$ & 0.0005 & 0.64 & 0.0008 & 1.60 & $-$0.0091$^{**}$ & $-$17.21 & $-$0.0114$^{**}$ & $-$14.88 \\ 
        $k = -5$ & 0.0009 & 1.80 & 0.0009$^{**}$ & 2.89 & $-$0.0059$^{**}$ & $-$17.72 & $-$0.0071$^{**}$ & $-$16.25 \\
\bottomrule
    \end{tabular}}
\end{table}

\restoregeometry

\end{document}